\documentclass[fleqn,usenatbib]{mnras}

\usepackage{newtxtext,newtxmath}

\usepackage[T1]{fontenc}
\usepackage{ae,aecompl}

\newcommand{\onefigsize}{\columnwidth}
\newcommand{\twofigsize}{\columnwidth}

\usepackage{amssymb}
\usepackage{amsmath}
\usepackage{color}
\usepackage{xcolor}
\usepackage{graphicx}
\usepackage{natbib}
\usepackage{times}
\usepackage[normalem]{ulem}

\title[Plasma Instabilities and TeV Blazar Spectra]{The Impact of Plasma Instabilities on the Spectra of TeV Blazars}

\author[R. Alves~Batista, A. Saveliev, E. M. de~Gouveia~Dal~Pino]{
Rafael Alves~Batista$^{1}$\thanks{Contact e-mail: \href{mailto:rafael.ab@usp.br}{rafael.ab@usp.br}},
Andrey Saveliev$^{2,3}$\thanks{Contact e-mail: \href{mailto:andrey.saveliev@desy.de}{andrey.saveliev@desy.de} (Corresponding author)},
Elisabete M. de~Gouveia~Dal~Pino$^{1}$\thanks{Contact e-mail: \href{mailto:dalpino@iag.usp.br}{adalpino@iag.usp.br}}
\\
$^{1}$Instituto de Astronomia, Geof{\'i}sica e Ci{\^e}ncias Atmosf{\'e}ricas, Universidade de S{\~a}o Paulo, 05508-090 S{\~a}o Paulo-SP, Brazil \\
$^{2}$Institute of Physics, Mathematics and Information Technology, Immanuel Kant Baltic Federal University, 236041 Kaliningrad, Russia\\
$^{3}$Faculty of Computational Mathematics and Cybernetics, Lomonosov Moscow State University, 119991 Moscow, Russia}

\date{Accepted 2019 August 26. Received 2019 August 23; in original form 2019 July 3}

\pubyear{2019}

\begin{document}

\label{firstpage}
\pagerange{\pageref{firstpage}--\pageref{lastpage}}
\maketitle

\volume{489}
\pagerange{3836-3849}

\begin{abstract}
Relativistic jets from blazars are known to be sources of very-high-energy gamma rays (VHEGRs). During their propagation in the intergalactic space, VHEGRs interact with pervasive cosmological photon fields such as the extragalactic background light (EBL) and the cosmic microwave background (CMB), producing electron-positron pairs. These pairs can upscatter CMB/EBL photons to high energies via inverse Compton (IC) scattering, thereby continuing the cascade process. This is often used to set limits on intergalactic magnetic fields (IGMFs). However, the picture may change if plasma instabilities, arising due to the interaction of the pairs with the intergalactic medium (IGM), cool down the electrons/positrons faster than inverse Compton scattering. As a consequence, the kinetic energy lost by the pairs to the IGM could cause a hardening in the observed gamma-ray spectrum at energies below $\sim$100 GeV. Here we study several types and models of instabilities and assess their impact for interpreting observations of distant blazars. Our results suggest that plasma instabilities can describe the spectra of some blazars and mimic the effects of IGMFs, depending on parameters such as the intrinsic spectrum of the object, the density and temperature of the IGM, and the luminosity of the beam. On the other hand, we find that for our fiducial set of parameters plasma instabilities do not have a major impact on the spectra of some of the blazars studied. Therefore, they may be used for constraining IGMFs.
\end{abstract}




\section{Introduction} \label{sec:Intro}

High-energy (i.e.~GeV and TeV) and very-high-energy gamma rays (VHEGRs) have become important tools to probe astrophysical and cosmological phenomena. This includes studies of our own galaxy~\citep{Weekes:1989tc,Strong:2004de,Aharonian:2005kn}, BL Lac objects \citep{1996ApJ...456L..83Q}, and gamma-ray bursts~\citep{Gehrels:2012kp}. Cosmological scales can also be probed with high-energy gamma rays, enabling constraints on the opacity of the Universe~\citep{Taylor:2011bn}, and heating of the intergalactic medium (IGM)~\citep{Chang:2011bf}, among others. The plethora of information that can be obtained with this messenger is possible mainly thanks to experimental efforts such as the High-Energy Stereoscopic System (H.E.S.S.)~\citep{Hofmann:1999ew}, the Very Energetic Radiation Imaging Telescope Array System (VERITAS)~\citep{Weekes:2001pd}, the Fermi Large Area Telescope (Fermi-LAT)~\citep{Atwood:2009ez}, and the Major Atmospheric Gamma Imaging Cherenkov (MAGIC)~\citep{Rico:2017euq}. Future air-Cherenkov telescopes such as the upcoming Cherenkov Telescope Array (CTA)~\citep{cta} will provide further insights into the high-energy universe.

Of particular interest are gamma-ray observations of TeV blazars, i.e.~active galactic nuclei (AGNs) whose jets are nearly parallel to the line of sight of emission. The standard model for the propagation of TeV gamma rays from the source to the observer is well-understood~\citep{PhysRevD.80.123012}. The emitted VHEGRs interact with background photons from pervasive radiation fields, predominantly the extragalactic background light (EBL), producing electron-positron pairs. The most relevant component of EBL in the context of this work is the infrared light, which consists mostly of diffuse photons stemming from star-formation processes, and starlight reprocessed by intergalactic dust~\citep{Ann.Rev.Astr.Astroph.39.1.249}. It is rather challenging to model the EBL \citep{Fan:2003qr}, especially at high redshifts. For this reason, a number of different models exists~\citep{Kneiske2004,Stecker:2005qs,Franceschini:2008tp,2010A&A...515A..19K,0004-637X-712-1-238,dominguez2011a,Gilmore:2011ks,stecker2016a}. This, in turn, results in different estimates of the mean free path of photons. In addition, higher order processes such as double pair production~\citep{Heiter:2017cev} and interactions with other pervasive photon fields like the radio background~\citep{Protheroe:1996si} are possible.
The pair-produced electrons (and positrons) can interact with low-energy photons from the cosmic microwave background (CMB), mostly via inverse Compton (IC) scattering. Here, the high-energy electrons upscatter the low-energy photons to energies in the GeV--TeV range. The mean free path of these electrons is typically $\lesssim 100 \; \text{kpc}$, much less than that for pair production ($\sim 100 \; \text{Mpc}$ at 10~TeV). Again, higher order processes like triplet pair production, even though highly suppressed, may be important for photons with energies $\gtrsim 10^{15} \; \text{eV}$~\citep{Heiter:2017cev}.


Intergalactic magnetic fields (IGMFs) can deflect the charged particles produced in the cascade,~i.e., electrons and positrons. Hence, the observed arrival direction of the secondary photons produced does not necessarily point back to the source. The result is that, instead of a point-like source (as would be the case in the absence of magnetic fields), a halo-like structure emerges~\citep{1994ApJ...423L...5A,JETPLett.85.10.473, PhysRevD.80.123012,0004-637X-703-1-1078,Biteau:2018tmv,Chen:2018mjd}. The size of the halo is determined by an interplay between the magnetic-field strength ($B$), its coherence length ($L_B$), and the electron energy. In general, $B$ relates to the deflection angle of electrons via the Lorentz force, whereas $L_B$ determines the propagation regime (diffusive when the Larmor radii of electrons is much smaller than the $L_B$, and ballistic otherwise). In addition, it has been shown that the shape of the halo is important as well as it may contain information about the magnetic helicity of IGMFs \citep{Tashiro:2013ita,AlvesBatista:2016urk,Duplessis:2017rde}, since helical fields produce spiral-like structures in the arrival directions.

The second effect of IGMFs on gamma-ray observables are time delays~\citep{Plaga1994,Fan:2003qr,1538-4357-686-2-L67,PhysRevD.80.123012,Fitoussi:2017ble}.  The effective propagation time of a secondary gamma ray which is observed as the result of the cascading processes described above is longer than the propagation time of a primary photon, i.e.~a gamma ray which arrives at Earth without interacting with the EBL. Therefore, if two such photons are emitted at the same time in a flare, they may be observed at different times, i.e.~the "reprocessed" photon arriving with a time delay with respect to the arrival time of the primary one.

Finally, another consequence of the propagation of VHEGRs in IGMFs is the suppression of the photon flux in the GeV range. This is a direct consequence of the energy dependence of the deflection angle. This angle increases dramatically with decreasing energy, such that GeV photons, indeed, are deflected away from the line of sight in such a way that they can not be observed at Earth, thus resulting in the aforementioned flux suppression. So far, this method has been used to derive only the lower limit on the IGMF strength~\citep{2002ApJ...580L...7D,d'Avezac:2007sg,Neronov02042010,Taylor:2011bn}.

While seemingly simple, the underlying mechanism of electromagnetic cascades is controversial. As first pointed out by \citet{0004-637X-752-1-22}, the pairs may interact with the IGM due to emerging plasma instabilities in the beam, cooling down before IC scattering can occur. This is a possible explanation for the absence of photons at GeV energies, observed in the spectra of some blazars. Yet, it should be noted that this interpretation, too, is disputed. While it is true that instabilities should occur when a relativistic beam traverses a background plasma, it is presently virtually impossible to directly simulate these scenarios, such that one has to rely on extrapolations which, however, are rather uncertain.

In this paper we study the spectra of some specific blazars for different plasma instability models. We first (section~\ref{sec:Mechanisms}) provide a brief description of the types of instabilities considered, including a discussion of the relevant physical parameters and conditions, accompanied by a number of models found in the literature. Then, in section~\ref{sec:SimSetup}, we describe the implementation of the instabilities in our Monte Carlo code. In section~\ref{sec:parameters} we study the effect of some relevant quantities on the expected gamma-ray fluxes for one specific scenario, while section~\ref{sec:blazars} gives an account of the predicted fluxes of gamma rays for a selected list of blazars. In section~\ref{sec:discussion} we discuss our results, and present our concluding remarks in section~\ref{sec:conclusions}.

\section{Plasma Instabilities in the IGM}\label{sec:Mechanisms}

Collective phenomena start to become relevant for high enough densities of electron-positron pairs. More precisely, the number of pairs within a spherical volume of radius equal to the wavelength of the most unstable mode -- the plasma skin depth -- which reads
\begin{equation}
    \lambda_\text{pl} \equiv \dfrac{2\pi c}{\omega_\text{pl}},
\end{equation}
where
\begin{equation}
    \omega_{\rm pl} = \sqrt{\frac{e^{2} n_{\rm IGM}}{\epsilon_{0} m_{e}}}
\end{equation}
gives the plasma frequency, and $n_\text{IGM}$ denotes the density of free-electrons in the IGM. 
In other words, plasma effects are relevant if the number of pairs in the beam within a volume of radius $\lambda_\text{pl}$ is much larger than unity, i.e., $\lambda_\text{pl}^3 n_\text{beam} \gg 1$, wherein $n_\text{beam}$ is the number density of pairs in the beam.

For the densities involved in blazar beams interacting with the IGM, $n_\text{beam} / n_\text{IGM} \ll 1$, thus ensuring that plasma effects are relevant for the evolution of the beams.

In this section we present some instabilities that might occur due to interactions between the beam and the background plasma. In principle, all of them can be relevant for the interaction between ultrarelativistic particles and the IGM.

\subsection{Types of instabilities}

The \emph{two-stream instability}, first described in \cite{Bohm:1949zz}, appears if the wave vector of an electrostatic perturbation (a Langmuir wave) is parallel to the flow for a range of wave numbers $0 < k \le \omega_{\rm pl}/v_{\rm beam}$. 

An intuitive picture of this instability can be obtained considering two flows moving towards each other. The lower-density flow -- the beam -- interacts with the higher-density background -- the plasma -- triggering the instability if the thermal velocity of the particles is smaller than the drift between the beams.
The Langmuir waves arising in the background plasma resonate with those in the beam, leading to the growth of the instability.

In a warm\footnote{A beam is considered to be warm if the condition $T \gtrsim \frac{3}{2^{10/3}} \left( \frac{n_{\rm beam}}{n_{\rm IGM}} \right)^{\frac{2}{3}} \gamma^{\frac{1}{3}} \frac{m_{e} c^{2}}{k_{\rm B}}$ is fulfilled, according to \citet{0004-637X-752-1-22}.} beam the velocity ($v_\text{beam}$) of some particles equals the phase velocity of the wave, hence being in resonance with it. Due to the form of the velocity distribution, more energy is transferred from the beam particles to the wave mode than vice-versa, thus resulting in an instability. For a cold-plasma beam the mechanism is slightly different, but still results in growing wave modes since the anisotropic electron and positron distribution functions interact with the co-moving background electrostatic wave, when the wave vector is parallel to the beam momentum, resulting in an instability growth time of \citep{0004-637X-752-1-22}
\begin{equation}
\mathcal{T} = \frac{2 \gamma}{\sqrt{3 n_{\rm IGM}}\omega_{\rm pl}} \left( \frac{n_{\rm IGM}}{n_{\rm beam}}\right)^{\frac{1}{3}}\,,
\end{equation}
where $n_{\rm IGM}$ and $n_{\rm beam}$ are the number densities of the IGM and the beam, respectively, $\gamma$ is the Lorentz factor of the particle and $\omega_{\rm pl}$ is the plasma frequency given by
\begin{equation}
\omega_{\rm pl} = \sqrt{\frac{e^{2} n_{\rm IGM}}{\epsilon_{0} m_{e}}}\,,
\end{equation}
with $\epsilon_{0}$ being the electrical vacuum permittivity, and $e$ and $m_{e}$ are the electron charge and mass, respectively.

The \emph{oblique instability} is, in principle, very similar to the two-stream instability described above. The main difference is the fact that in this case the wave vector is not parallel to the beam velocity; instead, the two are at an angle $\theta$ to each other. It can then be shown \citep{Nakar:2011mt} that in the astrophysical setting the dispersion relation is given by
\begin{equation}
1 =  \frac{n_{\rm beam}}{n_{\rm IGM}} \frac{m_{\rm beam}}{m_{\rm IGM}} \Psi\left( \Gamma, \gamma, \mathbf{k},\omega \right) - \left( \omega - C \frac{k c \cos\theta n_{\rm beam}}{\omega_{\rm pl}n_{\rm IGM}}\right)^{-1} \,,
\end{equation}
where $\mathbf{k}$ is the wave vector, $C$ is the fractional charge of the beam, $\Gamma$ is the Lorentz factor and $\Psi$ is a function of the given quantities. For $\theta = 0$ this equation can be solved analytically and gives the growth time for the two-stream instability discussed above. To find the growth time for the oblique instability at an arbitrary angle $\theta$ one has, in general, to resort to numerical simulations. As pointed out by \cite{0004-637X-752-1-22}, these two growth rates can differ significantly due to the fact that the relevant quantity for the beam--wave interaction is the projected velocity of the beam onto the wave vector, which for large $\theta$ is rather small, such that it is easier to create deflections and, subsequently, instabilities.

The \emph{modulation instability} in its simplest form can be explained as the result of ions in a turbulent medium scattering off oscillations caused by the beam. For this reason, they are transferred to smaller wave numbers, shifting the wave energy to higher (even superluminal) phase speeds \citep{GaleevJETP1977}. This takes place if the energy density of the electrostatic fluctuations $\epsilon_{e}$, given by \citep{Schlickeiser:2002dt}
\begin{equation}
\epsilon_{e} = n_{\rm beam} m_{e} c^{2} (\gamma - 1)
\end{equation}
becomes larger than the critical value \citep{1975FizPl...1...10G,Schlickeiser:2013eca}
\begin{equation}
\epsilon_{\rm crit} = \frac{5}{3} \frac{n_{\rm IGM} \left( k_{\rm B} T_{\rm IGM} \right)^{2}}{m_{e} c^{2}}\,.
\end{equation}
If this condition is fulfilled, one can find the growth rate for the modulation instability by analyzing the dispersion relation
\begin{equation}
1 = \frac{\omega_{\rm pl}^{2}}{\omega^{2}} + \frac{\omega^{2} \sin^{2}\theta \left( 1 - \frac{\gamma^{2} - 1}{\gamma^{2}} \cos^{2}\theta \right)}{\left( \omega - k c \cos\theta \sqrt{\frac{\gamma^{2} - 1}{\gamma^{2}}} \right)^{2}}\,.
\end{equation}
It should be noted here that formally the modulation instability is nothing but an oscillating two-stream instability \citep{1975PhFl...18.1769P,1976PhFl...19..849S,0004-637X-758-2-102}.

The \emph{Weibel instability}, named after the author of the seminal work on the topic, \cite{Weibel:1959zz}, is the result of the interaction of several counter-streaming beams \citep{1959PhFl....2..337F}. In its simplest form, following \cite{Medvedev:1999tu}, we can consider two beams in opposite directions with a vanishing net current. Small anisotropies in velocity cause fluctuations of the magnetic field perpendicular to the direction of the beam. For example, without loss of generality, if we take the particles to move along the $x$-axis, then the magnetic fluctuation can be represented in the form
\begin{equation}
\mathbf{B} = B  \cos\left( k y \right) \mathbf{e}_{z}\,.
\end{equation}
The resulting Lorentz force deflects the electrons in the beams in such a way that two current sheets are produced. Such large-scale currents induce larger magnetic fields which, in turn, increase the force, thus enhancing the effect and consequently producing the instability. The instability growth time is then given by \citep{0004-637X-752-1-22}
\begin{equation}
\mathcal{T} = \sqrt{\frac{n_{\rm beam} e^{2}}{\epsilon_{0} m_{e} \gamma}}\,.
\end{equation}
Note that while the Weibel instability does not contain oscillatory modes, they are present in the two-stream instability, which represents a fundamental difference between the two. 

The Weibel instability grows until the magnetic field is amplified to a point where the Larmor radii of the particles become comparable to the plasma skin depth, when the growth saturates~\citep{davidson1972a}. Moreover, small temperature gradients perpendicular to the beam direction can kill off this instability, sometimes giving it a short lifetime~\citep{0004-637X-752-1-22}.

Finally, another possible mechanism is \emph{non-linear Landau damping}. As the name suggests, the effect is highly non-linear and may be described as induced scattering by thermal ions, such that the frequency and wave vector of a given plasma (Langmuir) wave are transformed to different values \citep{0029-5515-14-6-012,Chang:2014cta}. The corresponding kinetic equation, which describes the wave transformation $(\mathbf{k},\omega) \rightarrow (\mathbf{k}',\omega')$, is given by
\begin{equation}
\frac{{\rm d} \epsilon_{\mathbf{k}}}{{\rm d}t} = 2 \frac{\epsilon_{\mathbf{k}}}{\mathcal{T}_{\mathbf{k}}} - \frac{\epsilon_{\mathbf{k}} \omega_{\rm pl}}{8 (2\pi)^{5/2} n_{e} m_{e} v_{e}^{2}} \int \frac{(\mathbf{k} \cdot \mathbf{k}')}{k^{2} k'^{2}} \phi(\mathbf{k},\mathbf{k}') \epsilon_{\mathbf{k}'} {\rm d}\mathbf{k}'\,,
\end{equation}
where $\epsilon_{\mathbf{k}}$ is the spectral energy density and $\phi(\mathbf{k},\mathbf{k}')$ is the overlap integral given in \cite{1968SvA....11..956K}. Here, we have $1/\mathcal{T}_{\mathbf{k}} = 1/\mathcal{T}_{\rm inst} + 1/\mathcal{T}_{\rm LD}$, where $\mathcal{T}_{\rm inst}$ denotes the instability growth time, and $\mathcal{T}_{\rm LD}$ denotes the growth time for linear Landau damping.

\subsection{Physical Parameters} \label{sec:PhysParam}

In this section we discuss the IGM and beam parameters which have an influence on the growth of plasma instabilities.

First, one should consider the \emph{number density of the IGM} ($n_{\rm IGM}$), which evolves as
\begin{equation} \label{nIGM}
n_{\rm IGM} = n_{\rm IGM,0} (1 + z)^{3}\,,
\end{equation}
wherein $z$ is the redshift at which the number density is calculated, and $n_\text{IGM,0}$ is the IGM density at present time. Our fiducial value for $n_{\rm IGM,0}$ is $n_\text{IGM,0} = 10^{-7} \; \text{cm}^{-3}$~\citep{0004-637X-752-1-22}. It should be noted that this is a rather simple model, not taking into account density fluctuations; these may be included if we multiply equation~\ref{nIGM} by $(1 + \delta)$, where $\delta$ is the overdensity.

The \emph{temperature of the IGM} ($T_{\rm IGM}$) has many uncertainties associated with it. Depending on whether one considers cosmic voids or the intracluster medium, the value of $T_{\rm IGM}$ ranges from $\sim 10^{3} \, {\rm K}$ \citep{Hui:1997dp} in voids, to $\sim 10^{9} \, {\rm K}$ in clusters~\citep{Reimer:2012vw,perucho2011a,burns2010a}. In order to account for this range, we adopt $T_{\rm IGM} = [10^{3}, 10^{4},10^{5},10^{6}]$, with $T_{\rm IGM} = 10^{4} \, {\rm K}$ being our fiducial value for the simulations presented below. This choice is fully justified if we consider that the volume filling factor of voids exceeds the one of clusters by orders of magnitude, so that the beam will interact predominantly with voids, hence the choice of temperatures. This range also encompasses the values considered by \citet{0004-637X-752-1-22}, \citet{Miniati:2012ge}, \citet{0004-637X-758-2-102}, \citet{Sironi:2013qfa}, and \citet{Vafin:2018kox}.


When it comes to the source parameters, the first important one is the \emph{total isotropic-equivalent luminosity} ($\mathcal{L}$). For TeV sources this value lies in the range $10^{41} \; \text{erg}\,\text{s}^{-1}$ and $10^{47}\; \text{erg}\,\text{s}^{-1}$ \citep{0004-637X-752-1-22}. We adopt the fiducial value of $\mathcal{L} = 10^{45} \; \text{erg}\,\text{s}^{-1}$. 

Another important quantity for the calculation of plasma instabilities is the \emph{beam plasma number density} ($n_{\rm beam}$). As this is a dynamic quantity, it is usually only possible to estimate it. In this work we follow \citet{0004-637X-752-1-22}, using an upper limit for $n_{\rm beam}$ which comes from analytical estimates of the cascade development assuming IC cooling and that the cooling is dominated by the kinetic oblique mode:
\begin{equation}
n_{\rm{beam}} \simeq 3.7 \times 10^{-22}\,{\rm cm^{-3}} \left( \dfrac{\mathcal{L}}{10^{45}\, \text{erg}\,\text{s}^{-1}} \right) \left( \dfrac{E_{e}}{10^{12}\,{\rm eV}} \right) \left( \dfrac{1+z}{2} \right)^{3 \zeta - 4}\,,
\label{eq:n_beam}
\end{equation}
wherein $\zeta = 4.5$ for $z<1$ is a parameter that can be inferred from the analysis of the local star formation rate \citep{Kneiske2004}.

\subsection{Models} \label{sec:Models}

Here we present and discuss the different models which we use to obtain blazar spectra in chronological order. In general, there are two different fundamental time scales to be considered. The first is the instability growth rate, which we label $\mathcal{T}_{i}$. The second is the energy loss time of the electrons/positrons due to plasma effects,~$\tau_{i}$, where in both cases $i$ refers to the model considered (see below). For the calculation of the blazar spectra the relevant quantity is the electron energy loss time $\tau_{i}$. In order to obtain the most robust lower limit for the photon flux, we adopt the estimate 
\begin{equation}
\tau_{i} = \mathcal{T}_{i}, 
\label{eq:taui}
\end{equation}
even though in principle $\tau_{i}$ might be substantially larger than $\mathcal{T}_{i}$~\citep{Grognard1975,2011PhPl...18d2307P}.

Note that while any of the instabilities described above can operate, for each of the models studied here there is always a dominant one. We summarise this in table~\ref{tab:models}. Further details are discussed below.

\begin{table*}[hbt!]
    \centering
    \caption{List of the models used and the instability that dominates in each case. We also identify the treatment employed in the original references.}
    \begin{tabular}{cccc}
        \hline \hline
        model & dominant instability & treatment & references \\
        \hline
        A & oblique & analytical &  \citet{0004-637X-752-1-22}  \\
        B & non-linear Landau damping & analytical &  \citet{Miniati:2012ge} \\
        C & modulation & analytical & \citet{0004-637X-758-2-102,Schlickeiser:2013eca} \\
        D & oblique & PIC &  \citet{Sironi:2013qfa} \\
        E & modulation & PIC & \citet{Vafin:2018kox} \\
        \hline
    \end{tabular}
    \label{tab:models}
\end{table*}

%

\subsubsection{Model A}

One of the first works on this topic and the one on which we mostly base our model A, is~\citet{0004-637X-752-1-22}, later on expanded and further analyzed by \cite{Chang:2011bf,Chang:2014cta,Chang:2016gji,Chang:2013yia,Shalaby:2017kpr,Shalaby:2018jja}. The authors distinguish between two regimes referred to as the `warm' and `cold' plasma beam cases depending on whether the beam density, $n_{\rm beam}$, is below (warm) or above (cold) the critical value $n_{\rm crit,A}$: 
\begin{equation}
\label{ncrit3} 
n_{\rm crit,A} = 1.6 \times 10^{-19} \left( \frac{E_{e}}{10^{12}\,{\rm eV}} \right)^{-2} \left( \frac{n_{\rm{IGM}}}{10^{-7}\,{\rm cm^{-3}}} \right) \;  {\rm cm^{-3}}.
\end{equation}
\cite{0004-637X-752-1-22} come to the conclusion that in both cases the oblique instability is the dominant one, however resulting in different expressions which can be summarized as
\begin{equation}
\tau_{\rm A} = 
\begin{cases}
\begin{split}
&7.1 \times 10^{7}\,{\rm s} \left( \dfrac{E_{e}}{10^{12}\,{\rm eV}} \right)^{-1} \left( \dfrac{n_{\rm{beam}}}{10^{-22}\,{\rm cm^{-3}}} \right)^{-1} \left( \dfrac{n_{\rm{IGM}}}{10^{-7}\,{\rm cm^{-3}}} \right)^{+\frac{1}{2}} \\ &\, \text{if} \, n_{\rm beam} \le n_{\rm crit,A} \, \text{(warm-plasma beam)} \,, \end{split}\\[10pt]
\begin{split}
&5.1 \times 10^{5}\,{\rm s} \left( \dfrac{E_{e}}{10^{12}\,{\rm eV}} \right)^{\frac{1}{3}}\left( \dfrac{n_{\rm{beam}}}{10^{-22}\,{\rm cm^{-3}}} \right)^{-\frac{1}{3}} \left( \dfrac{n_{\rm{IGM}}}{10^{-7}\,{\rm cm^{-3}}} \right)^{-\frac{1}{6}} \\ & \, \text{if} \, n_{\rm beam} > n_{\rm crit,A} \, \text{(cold-plasma beam)} \,,
\end{split}
\end{cases}
\label{eq:tau_A}
\end{equation}
For typical values ($n_\text{IGM} \sim 10^{-7} \; \text{cm}^{-3}$ and $n_\text{beam} \sim 10^{-22} \; \text{cm}^{-3}$) we find ourselves in a regime where the second expression from equation~\ref{eq:tau_A} holds, assuming that the electrons have energies $E_e \sim \; \text{a few TeV}$. If the electrons have higher energies ($E_e \sim 10 \; \text{TeV}$), then in overdense regions of the IGM $n_\text{beam} \gtrsim n_\text{crit,A}$, so that $\tau_A$ is given by the first expression of eq.~\ref{eq:tau_A}.

\subsubsection{Model B}

\cite{Miniati:2012ge} consider the interplay between Langmuir waves and non-linear Landau damping to be the most relevant effect. They have taken into account the finite nature of the spread of the momentum component transverse to the beam, which were obtained from Monte Carlo simulations of the cascade development.
They conclude that due to that the effect of plasma oscillations on energy losses of the electron-positron beams is mostly negligible, since inhomogeneities in the IGM can break the resonance between modes of the Langmuir and the beam waves.  Nevertheless, for the sake of completeness, we consider it here and compare it with the other models.

The authors obtained the energy loss time, $\tau_{\rm B}$, through a combination of Monte Carlo simulations to determine the beam properties as it propagates, and analytic solutions of appropriate equations, to obtain the instability growth rate. We can write
\begin{equation}
\tau_{\rm B} = \tau_{\rm IC} \mathfrak{T}_{\rm B}(D) (1 + z)^{2} = 3.9 \times 10^{13} \, {\rm s} \, (1 + z)^{-2} \left( \frac{E_{e}}{10^{12} \, {\rm eV}} \right)^{-1} \mathfrak{T}_{\rm B}(D)\,,
\label{eq:tau_B}
\end{equation}
where $E_{e}$ is the electron/positron energy, $D$ is the co-moving distance to the source, and $\mathfrak{T}_{\rm B}(D)$ is tabulated as shown in table~\ref{TableMiniati}.

\begin{table*}
\centering
\caption{Tabulated values for $\mathfrak{T}_{\rm B}(D)$ from equation~\ref{eq:tau_B}.}
\begin{tabular}{|c|cccccccccccccccc|}
\hline 
$\log_{10}(D/{\rm Mpc)}$ & -0.05 & 0.14 & 0.35 & 0.59 & 0.79 & 0.96 & 1.17 & 1.40 & 1.57 & 1.77 & 1.99 & 2.20 & 2.41 & 2.60 & 2.80 & 3.00 \\
\hline
$\log_{10}\mathfrak{T}_{\rm B}(D)$ & 3.28 & 3.46 & 3.14 & 3.08 & 3.05 & 3.00 & 2.84 & 2.56 & 2.40 & 2.55 & 2.36 & 2.08 & 1.73 & 1.55 & 1.05 & 0.75 \\
\hline
\end{tabular} 
\label{TableMiniati}
\end{table*}

From table~\ref{TableMiniati} one can see that non-linear Landau damping is more important closer to the blazar. On the other hand, the inhomogeneities in the IGM are more relevant at larger distances. The interplay between these two effects lead to an effective stabilisation of them beam, thus explaining why the instabilities are completely subdominant in this scenario.

\subsubsection{Model C}

This model is based on analytic calculations developed by \cite{0004-637X-758-2-102,Schlickeiser:2013eca}. In particular, in \cite{0004-637X-758-2-102} the authors find that the different effects contribute to the suppression of the electromagnetic cascade at different regimes. In the `strong  blazar' regime, the beam plasma density is lower than a certain critical value $n_{\rm crit,C}$, resulting in a dominance of the modulation instability. If $n_{\rm beam} > n_{\rm crit,C}$ -- the `weak blazar' case -- the modulation instability does not set in. In this case, non-Linear Landau damping becomes relevant, such that energy is deposited in electrostatic and electromagnetic fluctuations in the background plasma. 

The energy loss time is
%
\begin{equation}
\tau_{\rm C} = 
\begin{cases}
\begin{split}
 &5.2 \times 10^{14}\,{\rm s} \left( \dfrac{E_{e}}{10^{12}\,{\rm eV}} \right)^{\frac{5}{3}}\left( \dfrac{n_{\rm{beam}}}{10^{-22}\,{\rm cm^{-3}}} \right)^{\frac{1}{3}} \left( \dfrac{n_{\rm{IGM}}}{10^{-7}\,{\rm cm^{-3}}} \right)^{-\frac{5}{6}} \\ &\times \left( \dfrac{T_{\rm IGM}}{10^{4}\,{\rm K}} \right)^{-2} \text{if} \, n_{\rm beam} \le n_{\rm crit,C} \, \text{(weak blazar)} \,,
 \end{split}
 \\
 \begin{split}
 &8.3 \times 10^{6}\,{\rm s} \left( \dfrac{E_{e}}{10^{12}\,{\rm eV}} \right)^{\frac{1}{3}} \left( \dfrac{n_{\rm{beam}}}{10^{-22}\,{\rm cm^{-3}}} \right)^{-\frac{1}{3}} \left( \dfrac{n_{\rm{IGM}}}{10^{-7}\,{\rm cm^{-3}}} \right)^{-\frac{1}{6}} \\ &\times \mathfrak{T}_{\rm C}(n_{\rm IGM},T_{\rm IGM}) \, \text{if} \, n_{\rm beam} > n_{\rm crit,C} \, \text{(strong blazar)} \,,
 \end{split}
\end{cases}
\label{eq:tau_C}
\end{equation}
where $\mathfrak{T}_{\rm C}$ is
\begin{equation}
\mathfrak{T}_{\rm C}(n_{\rm IGM},T_{\rm IGM}) = 1 + \frac{5}{4} \ln\left( \frac{T_{\rm IGM}}{10^{4}\,{\rm K}} \right) - \frac{1}{4} \ln\left( \frac{n_{\rm{IGM}}}{10^{7}\,{\rm cm^{-3}}} \right)
\end{equation}
and $n_{\rm crit,C}$ is given by
\begin{equation}
\label{ncrit2} 
n_{\rm crit,C} = 2.5 \times 10^{-25} \left( \dfrac{E_{e}}{10^{12}\,{\rm eV}} \right)^{-1} \left( \dfrac{n_{\rm{IGM}}}{10^{-7}\,{\rm cm^{-3}}} \right) \left( \dfrac{T}{10^{4}\,{\rm K}} \right)^{2} \;  {\rm cm^{-3}} \, .
\end{equation}

Note that while the cooling rate of electrons depends quadratically on the temperature in the weak blazar case, in the strong blazar case this dependence is much weaker. This is a consequence of the onset of the modulation instability, which depends on both the beam density, $n_\text{beam}$, as well as on the IGMF temperature, $n_\text{IGM}$.

\subsubsection{Model D}

This model is based on \cite{Sironi:2013qfa}. The authors distinguish two cases, the `cold-plasma beam' and the `warm-plasma beam', depending on whether the beam plasma density $n_{\rm beam}$ is above or below value $n_{\rm crit,D}$, respectively. For both cases they find that the oblique instability is the most relevant one, resulting in the energy loss time
\begin{equation}
\tau_{\rm D} = 
\begin{cases}
\begin{split}
&1.4 \times 10^{7}\,{\rm s} \left( \dfrac{E_{e}}{10^{12}\,{\rm eV}} \right)^{-1} \left( \dfrac{n_{\rm{beam}}}{10^{-22}\,{\rm cm^{-3}}} \right)^{-1} \left( \dfrac{n_{\rm{IGM}}}{10^{-7}\,{\rm cm^{-3}}} \right)^{+\frac{1}{2}} \\
&\,\text{if} \, n_{\rm beam} \le n_{\rm crit,D} \, \text{(warm-plasma beam)} \,, \end{split}
\\
9.6 \times 10^{5}\,{\rm s} \left( \dfrac{E_{e}}{10^{12}\,{\rm eV}} \right)^{\frac{1}{3}}\left( \dfrac{n_{\rm{beam}}}{10^{-22}\,{\rm cm^{-3}}} \right)^{-\frac{1}{3}} \left( \dfrac{n_{\rm{IGM}}}{10^{-7}\,{\rm cm^{-3}}} \right)^{-\frac{1}{6}} \\
\, \text{if} \,n_{\rm beam} > n_{\rm crit,D} \, \text{(cold-plasma beam)} \,,
\end{cases}
\label{eq:tau_D}
\end{equation}
where $n_{\rm crit,D}$ is given by
\begin{equation}
\label{ncrit1} 
n_{\rm crit,D} = 8.0 \times 10^{-20}  \left( \dfrac{E_{e}}{10^{12}\,{\rm eV}} \right)^{-2} \left( \dfrac{n_{\rm{IGM}}}{10^{-7}\,{\rm cm^{-3}}} \right) \; {\rm cm^{-3}} \,.
\end{equation}

At first glance, this model is remarkably similar to model~A, apart from some numerical factors (compare equations~\ref{eq:tau_A} and~\ref{eq:tau_D}). However, these two models are intrinsically different. While \citet{0004-637X-752-1-22} argued that the background plasma takes about $\sim 50\%$ of the beam energy as the beam relaxation time is much shorter than IC losses, the PIC simulations of \citet{Sironi:2013qfa} result in an energy-transfer factor of $\sim 10^{-1}$. This discrepancy stems, in part, from the treatment of the transverse momentum, which inherits from the momentum distribution of the pairs generated in the cascade. 

\citet{Sironi:2013qfa} also note that the oblique exponential growth takes place on timescales much shorter than the time it takes for the relaxation phase to be reached. For this reason, there is an associated efficiency factor that indicates how much of the beam's energy is used to heat the IGM, which depends on the spectrum of the component of the electrons' momenta perpendicular to the beam. Here we take this factor to be 1, since we are interested in investigating the scenario that provides the \emph{maximum} quenching of the cascade.  

\subsubsection{Model E}

The final model we adopt is mostly taken from \cite{Vafin:2018kox}, who find that the modulation instability dominates the energy losses. Through a thorough analysis, they arrive at
\begin{equation}
\begin{split}
\tau_{\rm E} &= 1.9 \times 10^{11} \, {\rm s} \left( \dfrac{E_{e}}{10^{12}\,{\rm eV}} \right)^{\frac{4}{3}}\left( \dfrac{n_{\rm{beam}}}{10^{-22}\,{\rm cm^{-3}}} \right)^{\frac{1}{3}} \left( \dfrac{n_{\rm{IGM}}}{10^{-7}\,{\rm cm^{-3}}} \right)^{-\frac{1}{3}} \\
&\times \left( \dfrac{T_{\rm IGM}}{10^{4}\,{\rm K}} \right)^{-1}\,.
\end{split}
\label{eq:tau_E}
\end{equation}
%
%
\citet{Vafin:2018kox} claim that non-linear Landau damping is properly resolved in their PIC simulations; in this case, the peak frequency is $\sim 10^{-5} \omega_\text{pl}$, whereas the peak frequency for the linear growth is $\sim 10^{-3} \omega_\text{pl}$. 

The electrostatic mode in stabilised by the modulation instability when the energy density stored in the waves is a fraction of that in the beam. This is related to an efficiency factor which, once again, we have assumed to be maximal for the purposes of this study. 

\section{Simulation Setup} \label{sec:SimSetup}

For this analysis we employ the Monte Carlo code CRPropa 3~\citep{Batista:2016yrx}. It enables the propagation of high-energy particles in the intergalactic space. It includes the relevant interactions for the development of electromagnetic cascades, namely inverse Compton scattering, pair production, and adiabatic losses due to the expansion of the Universe. We consider the CMB and the EBL as target background fields and choose the EBL model by \citet{Gilmore:2011ks} for the latter. This choice quantitatively affects the shape of the simulated spectra, but it is virtually irrelevant for the qualitative discussion that follows.

\begin{figure}
\centering
\includegraphics[width=\onefigsize]{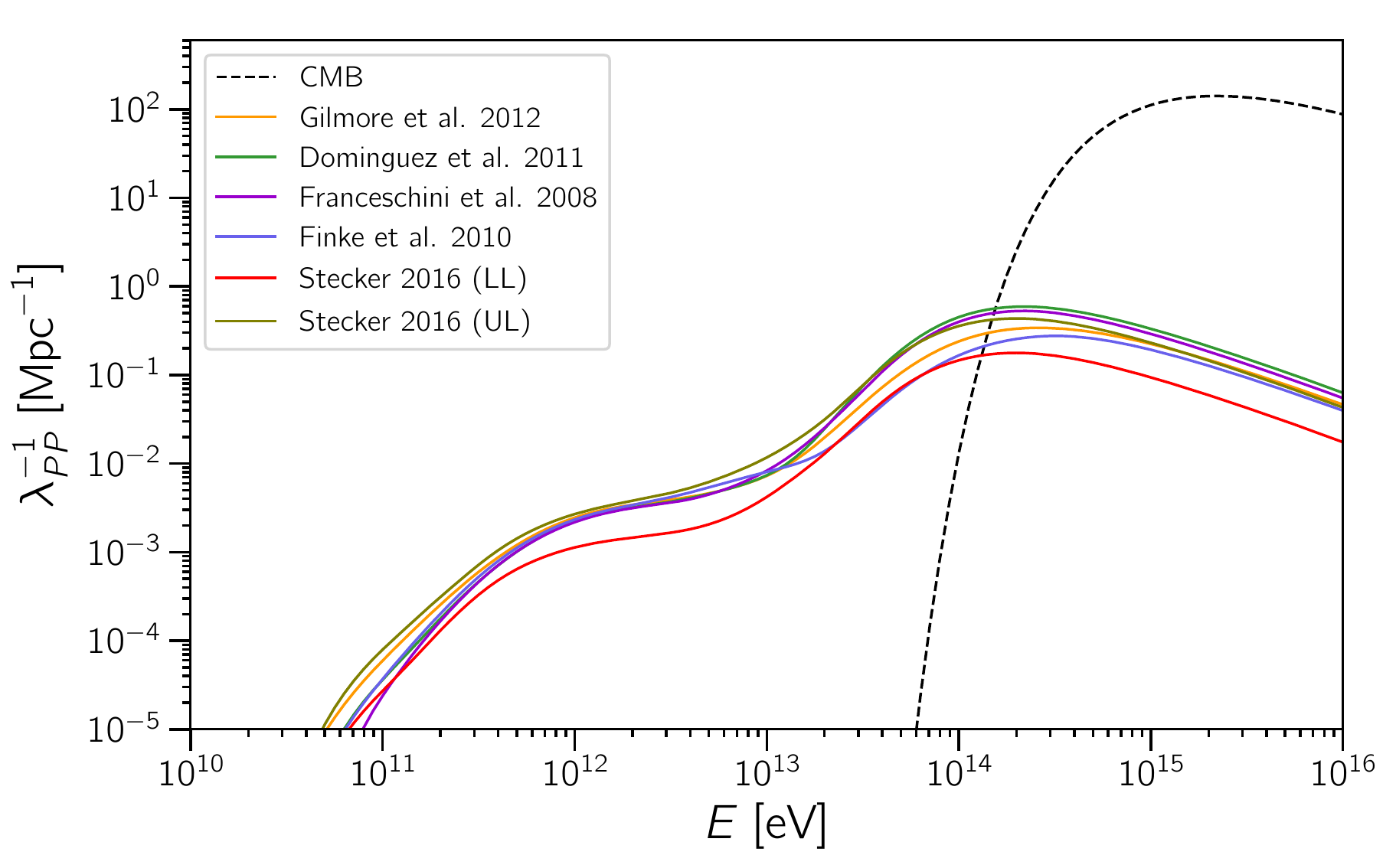}
\caption{Interaction rates for pair production computed at $z=0$ for several EBL models (solid lines) and the CMB (dashed line). The EBL models considered are: \citet{Gilmore:2011ks}, \citet{dominguez2011a}, \citet{Franceschini:2008tp}, \citet{0004-637X-712-1-238}, and the upper (UL) and lower (LL) limit models by \citet{stecker2016a}.}
\label{fig:ratePP}
\end{figure}

The most important background for pair production at the energies of interest ($E \lesssim 10^{14} \; \text{eV}$) is the EBL, whose interaction rates are shown in figure~\ref{fig:ratePP} for a few different models. Although this figure provides a detailed model for pair production, a rough analytical estimate of the mean free path for this process yields~\citep{PhysRevD.80.123012}
\begin{equation}
\lambda_\text{PP} \simeq 80\;{\rm Mpc} \frac{\kappa}{(1 + z)^{2}}\left( \frac{E_{\gamma}}{10^{13}\;{\rm eV}} \right)^{-1}\,,
\end{equation}
where $E_{\gamma}$ is the photon energy, $z$ is the redshift and $\kappa$ is an energy- and redshift-dependent coefficient to account for uncertainties in the EBL.

For electrons, the dominant background is the CMB. The energy-loss length, in this case, is nicely approximated by~\citep{PhysRevD.80.123012}
\begin{equation}
\lambda_\text{IC} \simeq 1.2 \; {\rm kpc} \, (1 + z)^{-3} \,.
\end{equation}
This approximate analytical expression agrees exceedingly well with what we obtain with our detailed Monte Carlo treatment. Note that $\lambda_\text{IC} \ll \lambda_\text{PP}$, implying that the charged component of the cascade does not live very long compared to photons.

We took advantage of the modular structure of CRPropa code to develop an extension\footnote{\url{https://github.com/rafaelab/grplinst}} dedicated to calculate the effects of plasma instabilities in an approximate fashion. To this end, we convert the energy loss times given by equations~\ref{eq:tau_A},~\ref{eq:tau_B},~\ref{eq:tau_C},~\ref{eq:tau_D}, and~\ref{eq:tau_E} to the energy loss per distance, $P(E,x,z)$, which in the ultrarelativistic limit can be written as
\begin{equation} \label{Ploss}
P(E_{e},x,z) \equiv - \frac{{\rm d}E_{e}}{{\rm d}x}(E_{e},x,z) = \frac{E_{e}}{c \tau(E_{e},x,z)}\,,
\end{equation}
where $x$ is the trajectory length propagated by a particle and $c$ is the speed of light.

We should stress that the values of $\tau$ for models B, C, and E are the actual energy loss times of the beams, as in our treatment; for models A and D, they correspond to the maximum instability growth time. As has been argued in the corresponding publications, the actual energy loss (and hence the value of $P$) can be significantly smaller. Therefore, all values of $\tau$ presented here should be interpreted as lower limits; this should be kept in mind for the analyses carried out in sections~\ref{sec:parameters} and~\ref{sec:blazars}.

To calculate the energy loss function $P$ for the different models, we combine the respective values for $\tau$ from section~\ref{sec:Models} with the model of the beam density $n_{\rm beam}$ from equation~\ref{eq:n_beam} (with $\zeta = 4.5$), as we are only interested in blazars with $z \ll 1.0$ and the value of $n_{\rm IGM}$ from equation~\ref{nIGM}.

\begin{figure}
\centering
\includegraphics[width=\onefigsize]{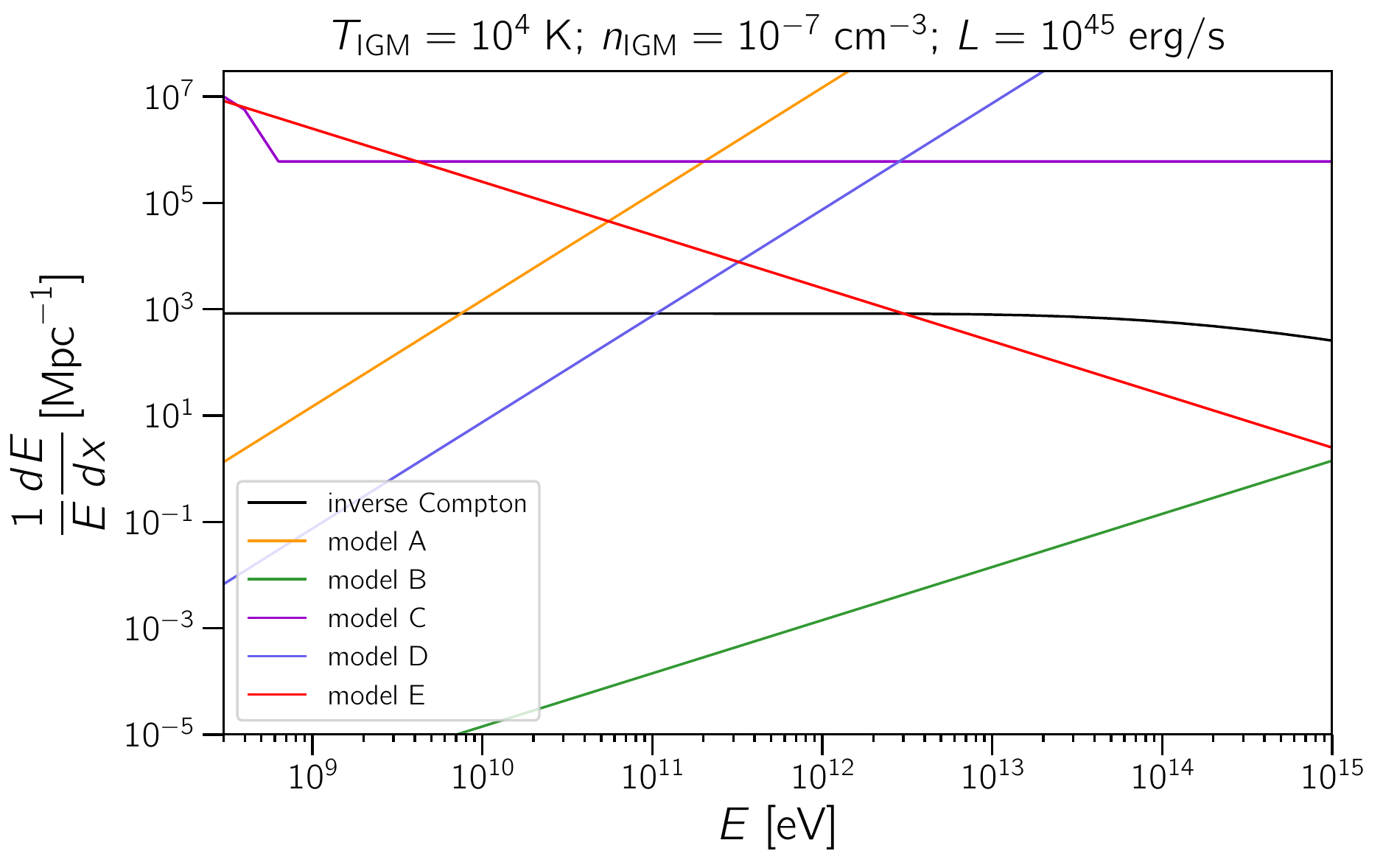}
\caption{Cooling rates computed at $z=0$ for the different models and a typical combination of parameters: $T_{\text{IGM},0} = 10^4 \; \text{K}$, $n_{\text{IGM},0}=10^{-7} \; \text{cm}^{-3}$, and $L_\text{beam}=10^{45} \; \text{erg/s}$. See table~\ref{tab:models} for a more details about the electron-plasma instability models considered. For reference, we also present the inverse mean free path for inverse Compton scattering.  
}
\label{fig:rate}
\end{figure}

In figure~\ref{fig:rate} we compare the electron cooling rates for the different models, for a typical combination of parameters. We also show the interaction rate for inverse Compton scattering. Note that the energy dependence across the different models differs considerably.


As a benchmark scenario, we consider the case of a source located at $z \approx 0.14$, which is approximately the redshift of the extreme blazar 1ES 0229+200, commonly used in studies of IGMFs~\citep{Neronov02042010,Vovk:2011aa,Arlen:2012iy,Yang:2015lqy,Yan:2018pca}. We model the spectrum injected by the blazar as a power law of the form
\begin{equation}
\dfrac{dN}{dE} \propto E^{-\alpha} \exp\left( -\dfrac{E}{E_\text{max}} \right).
\end{equation}
Unless stated otherwise, we adopt the parameters from~\citet{Vovk:2011aa}: $\alpha=1.2$ and $E_\text{max}=5\times 10^{12} \; \text{eV}$. We use the EBL model by \citet{Gilmore:2011ks}, although this choice should not significantly change the interpretation of the results.

\section{Simulation results}
\label{sec:parameters}

Now we present the results of the simulations. Throughout this section, we present the results for our benchmark scenario, fixing $T_{\text{IGM},0}=10^{4} \; \text{K}$, $n_{\text{IGM},0}=10^{-7} \; \text{cm}^{-3}$, and $\mathcal{L}=10^{45} \; \text{erg/s}$ and varying one of the parameters at a time. As stated in section~\ref{sec:PhysParam}, we also fix $\alpha=1.2$ and $E_\text{max}=5 \; \text{TeV}$.

\subsection{Plasma instability models}

\begin{figure}
  \centering
  \includegraphics[width=\onefigsize]{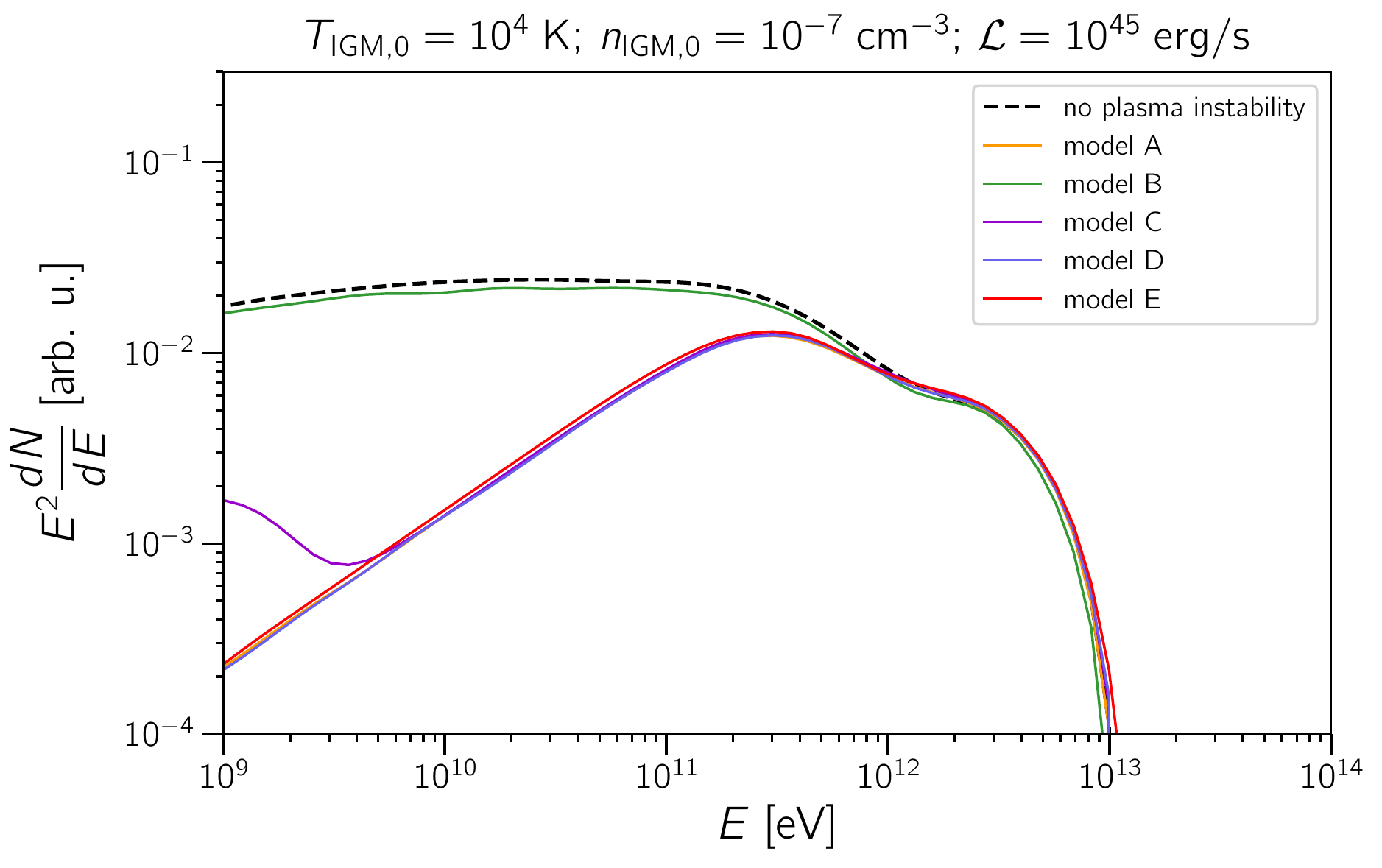}
  \caption{Gamma-ray fluxes observed at Earth for the benchmark scenario and the different models, assuming $n_{\text{IGM},0}=10^{-7} \; \text{cm}^{-3}$ and $\mathcal{L}=10^{45} \; \text{erg/s}$. The scenario with no plasma instabilities is represented as a dashed line. See table~\ref{tab:models} for a more details about the plasma instability models considered.}
  \label{fig:models}
\end{figure}

We first compare all the models for our benchmark scenario. This is shown in figure~\ref{fig:models}. Note that the slope for all models but B are rather similar in the energy range of $\sim 10 - 300 \; \text{GeV}$. This is expected because the spectral index of injection ($\alpha = 1.2$ for the benchmark scenario) should be retrieved if virtually all secondary electrons are removed from the cascade. If plasma instabilities are not as efficient as we considered, then the picture would change and a fraction of the electrons would remain in the cascade. If identified, this slope would be a phenomenological signature of a very efficient process quenching the cascade; plasma instabilities would then be a candidate explanation. Departures from this behaviour may occur for different values of $n_{\text{IGM},0}$, $T_{\text{IGM},0}$, and $\mathcal{L}$, as will be discussed later. Nevertheless, for $30 \lesssim E / \text{GeV} \lesssim 100$ all models A, C, D, and E led to a slope of $E^{-1.2}$ at energies $\lesssim 100 \; \text{GeV}$ in most cases studied, the exception being model C for some particular combinations of parameters. 

We confirm the results by~\citet{Miniati:2012ge}, who argue that plasma effects do not lead to noticeable changes in the observed gamma-ray flux. This is shown in figure~\ref{fig:models} and can be confirmed by analysing figure~\ref{fig:rate}, which shows that cooling rates for model B are orders of magnitude below the interaction rate for IC scattering at the energy range of interest ($E \lesssim 1 \; \text{PeV}$). For this reason, we omit plots for model B when studying the effects of temperature, IGM density, beam luminosity, and spectral parameters.

\subsection{Temperature of the IGM}

We study the effects of the temperature on the spectrum for our benchmark scenario. We vary only the temperature, while keeping all other quantities ($n_{\text{IGM},0}=10^{-7} \; \text{cm}^{-3}$ and $\mathcal{L}=10^{45} \; \text{erg/s}$) fixed.

\begin{figure*}
  \centering
  \includegraphics[width=\twofigsize]{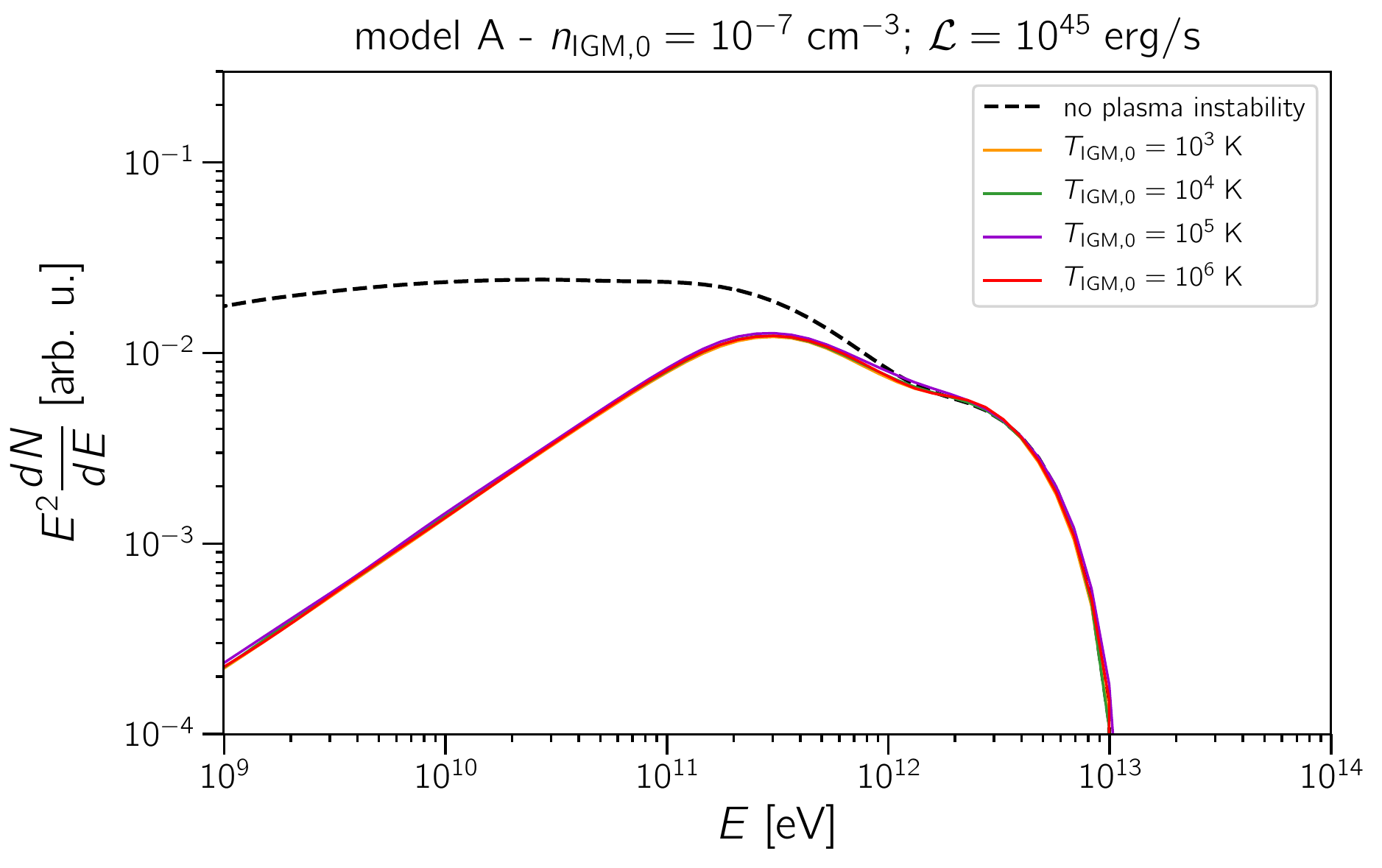}
  \includegraphics[width=\twofigsize]{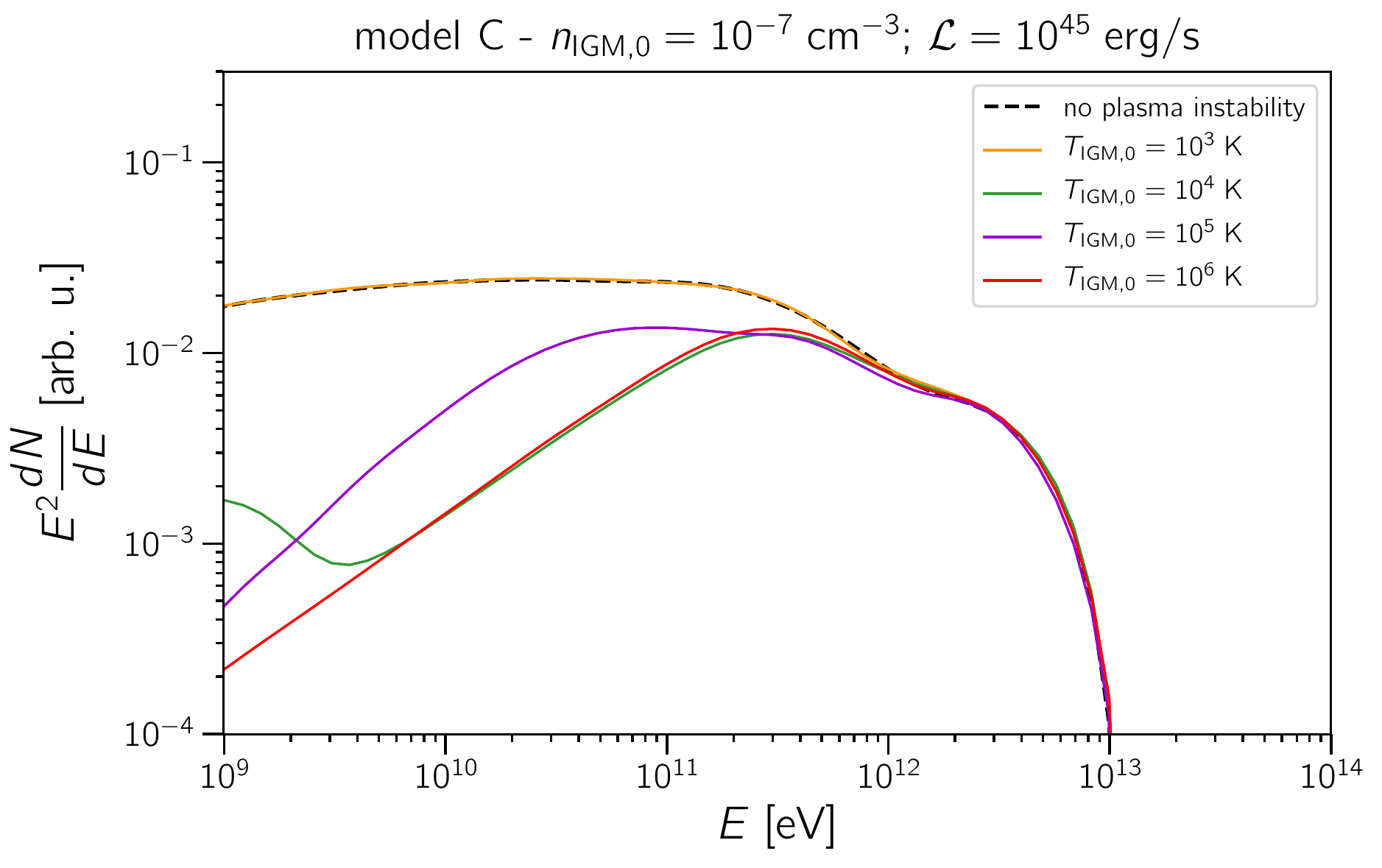}
  \includegraphics[width=\twofigsize]{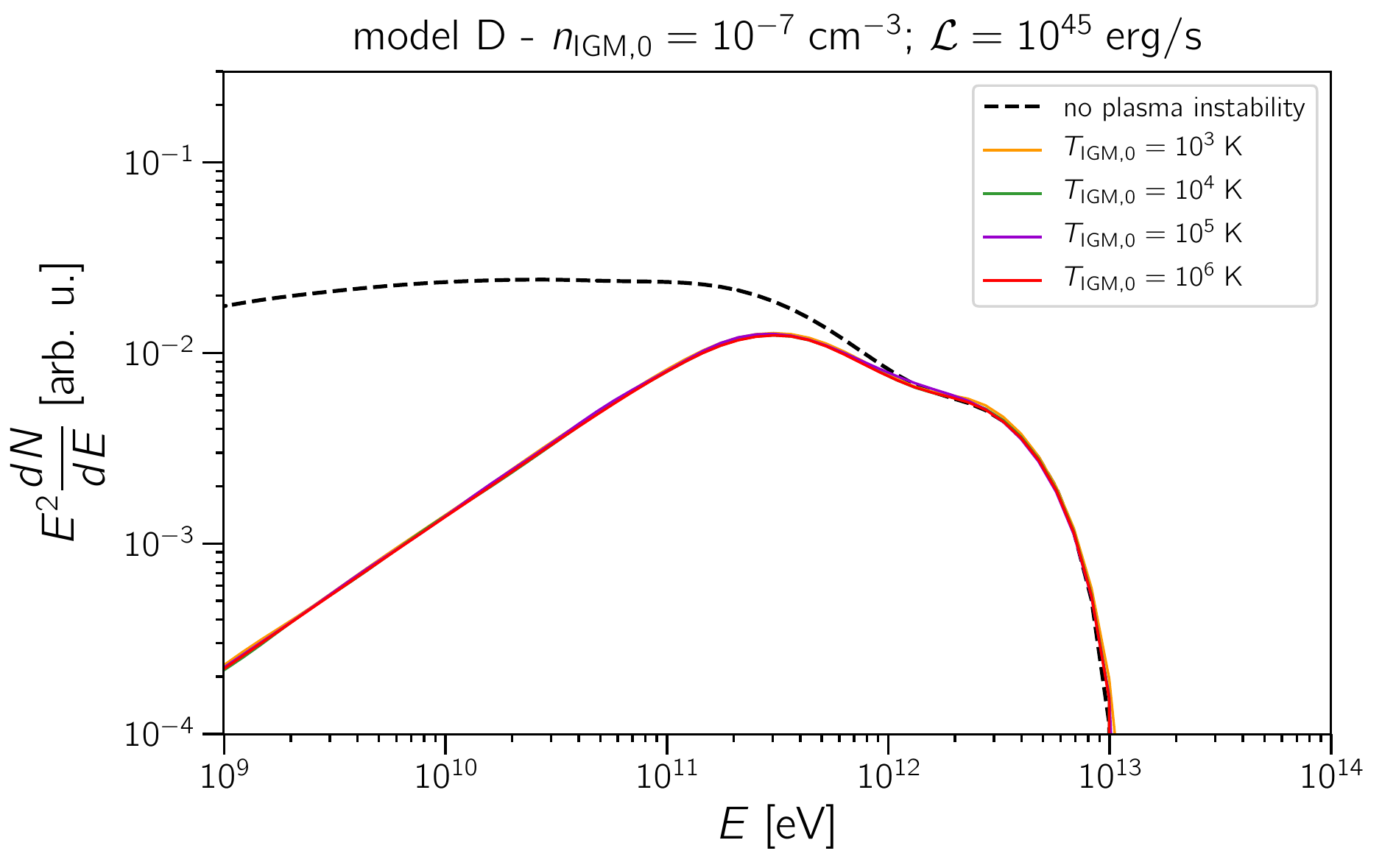}
  \includegraphics[width=\twofigsize]{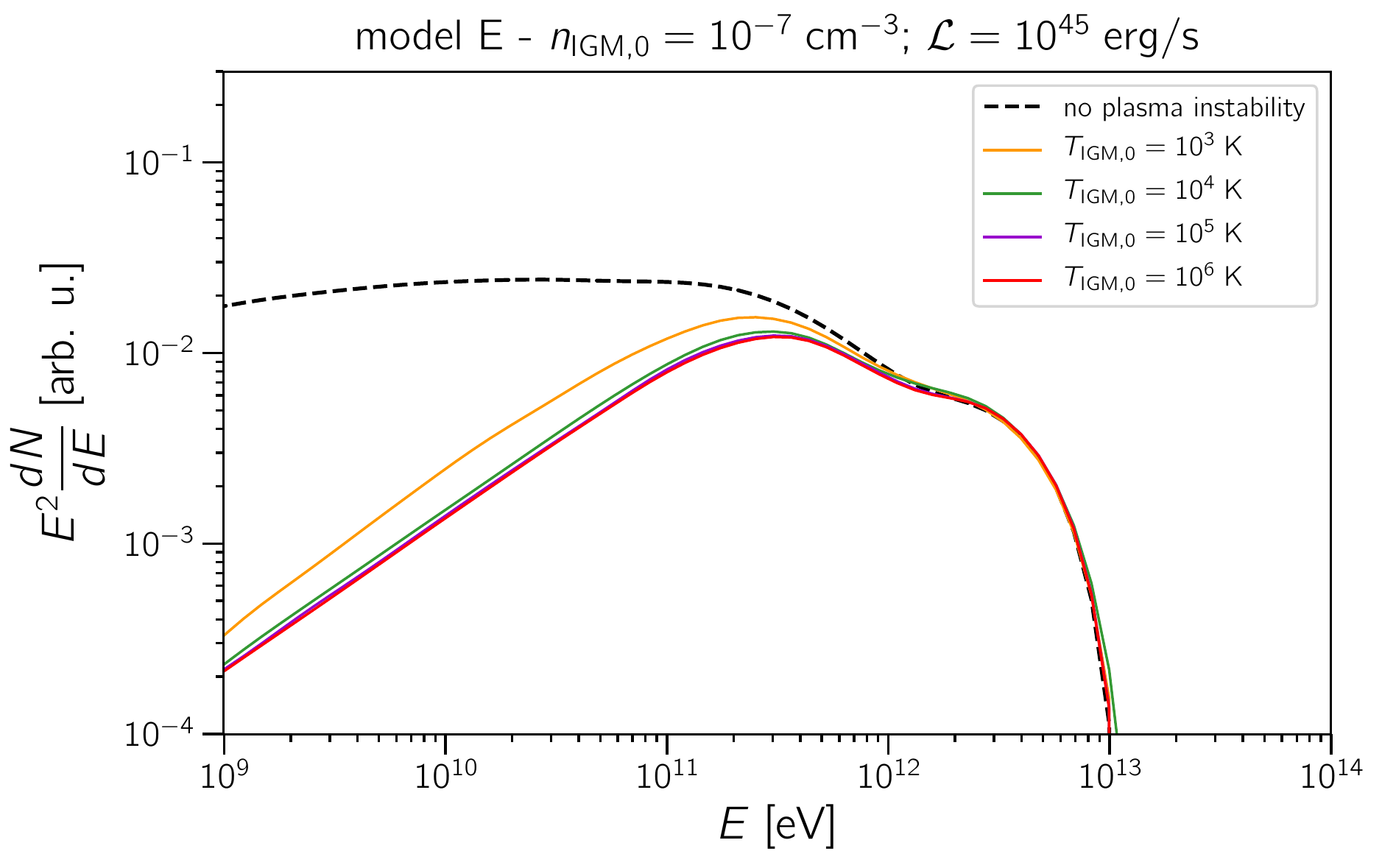}
  \caption{Gamma-ray fluxes observed at Earth for the benchmark scenario, assuming $n_{\text{IGM},0}=10^{-7} \; \text{cm}^{-3}$ and $\mathcal{L}=10^{45} \; \text{erg/s}$. }
  \label{fig:temperature}
\end{figure*}

As shown in figure~\ref{fig:temperature}, the temperature dependence is more prominent for models C and E, while models A and D are virtually temperature-independent. This follows immediately from equations~\ref{eq:tau_A}, \ref{eq:tau_C}, \ref{eq:tau_D}, and \ref{eq:tau_E} as there are no explicit temperature dependencies in equations~\ref{eq:tau_A} and \ref{eq:tau_D}.

At $T_\text{IGM,0} \sim 10^{4} \; \text{K}$ the spectrum for model C presents a change in slope at energies $\lesssim 3 \; \text{GeV}$. This may be an observational signature of this model, and if observed in multiple blazars, would allow us to draw conclusions about the temperature or correlated parameters.

\subsection{Density of the IGM}

\begin{figure*}
  \centering
  \includegraphics[width=\twofigsize]{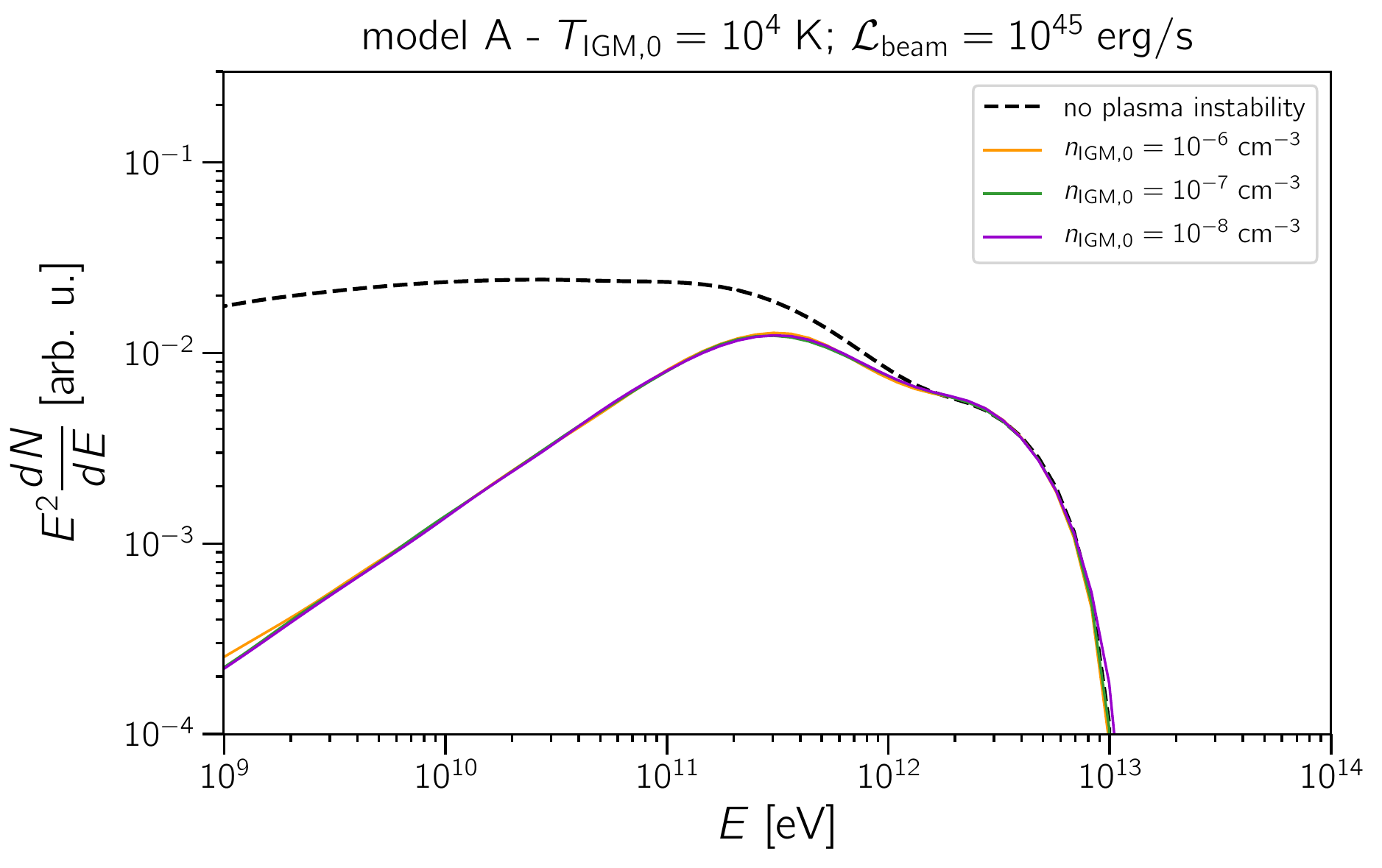}
  \includegraphics[width=\twofigsize]{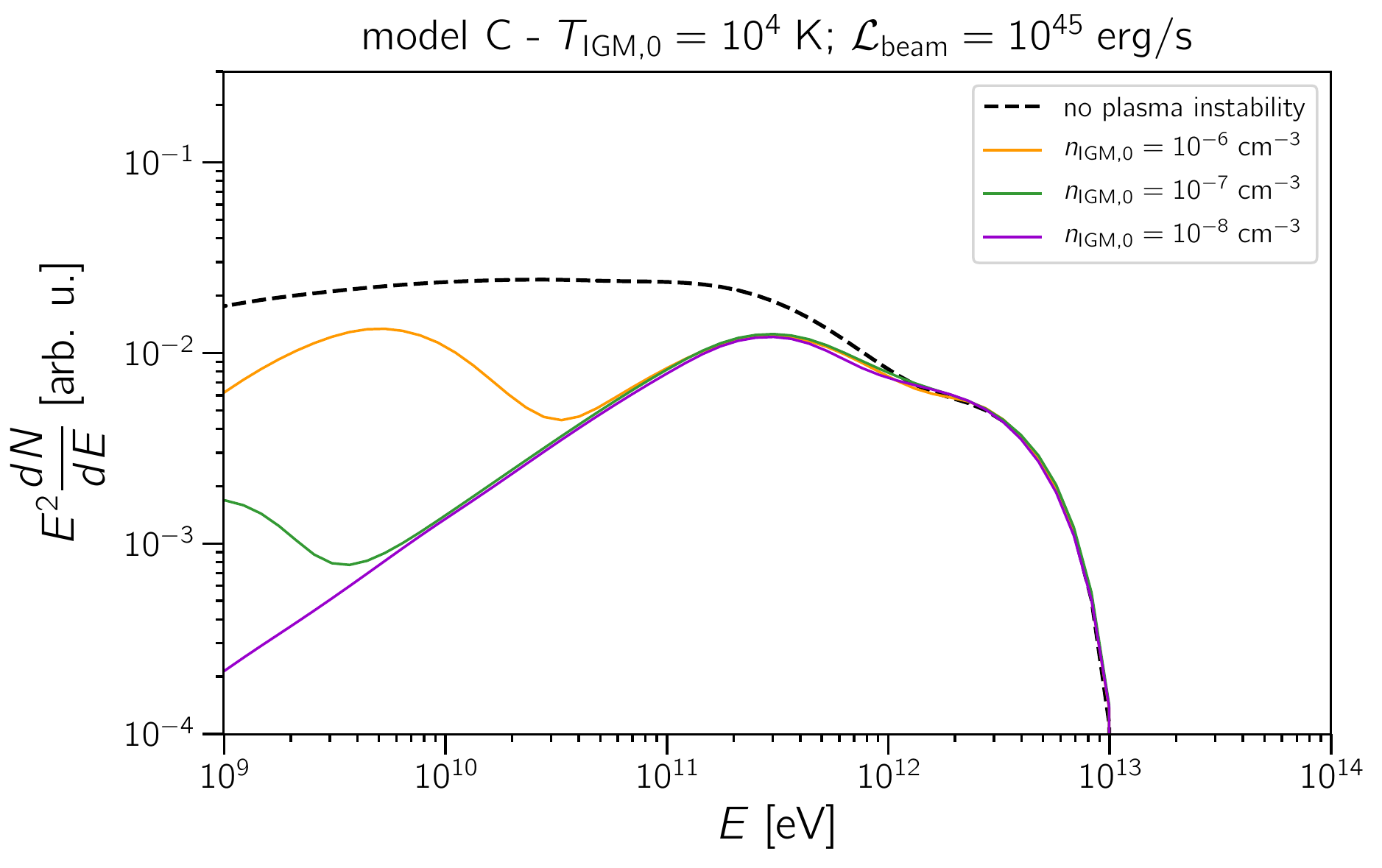}
  \includegraphics[width=\twofigsize]{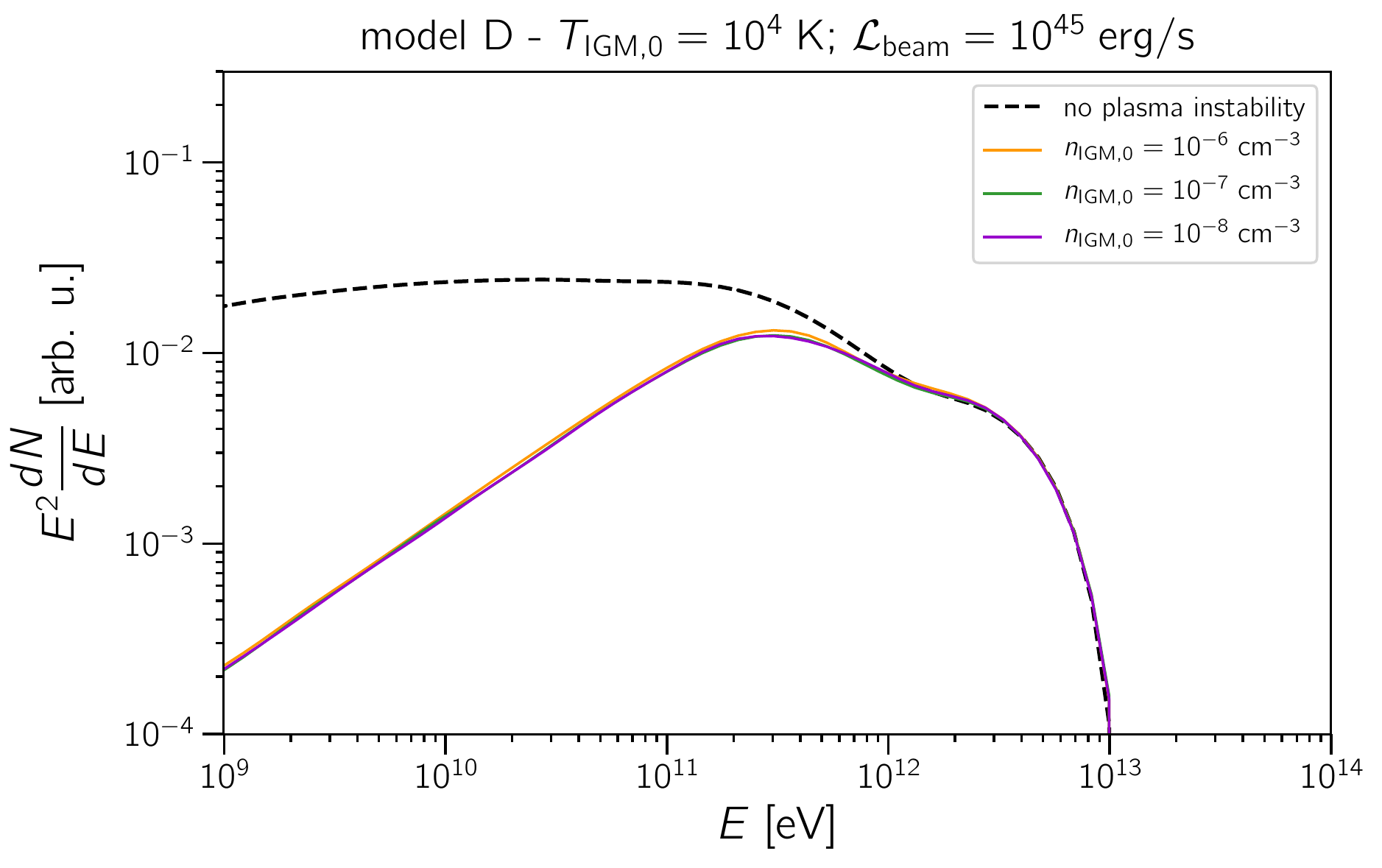}
  \includegraphics[width=\twofigsize]{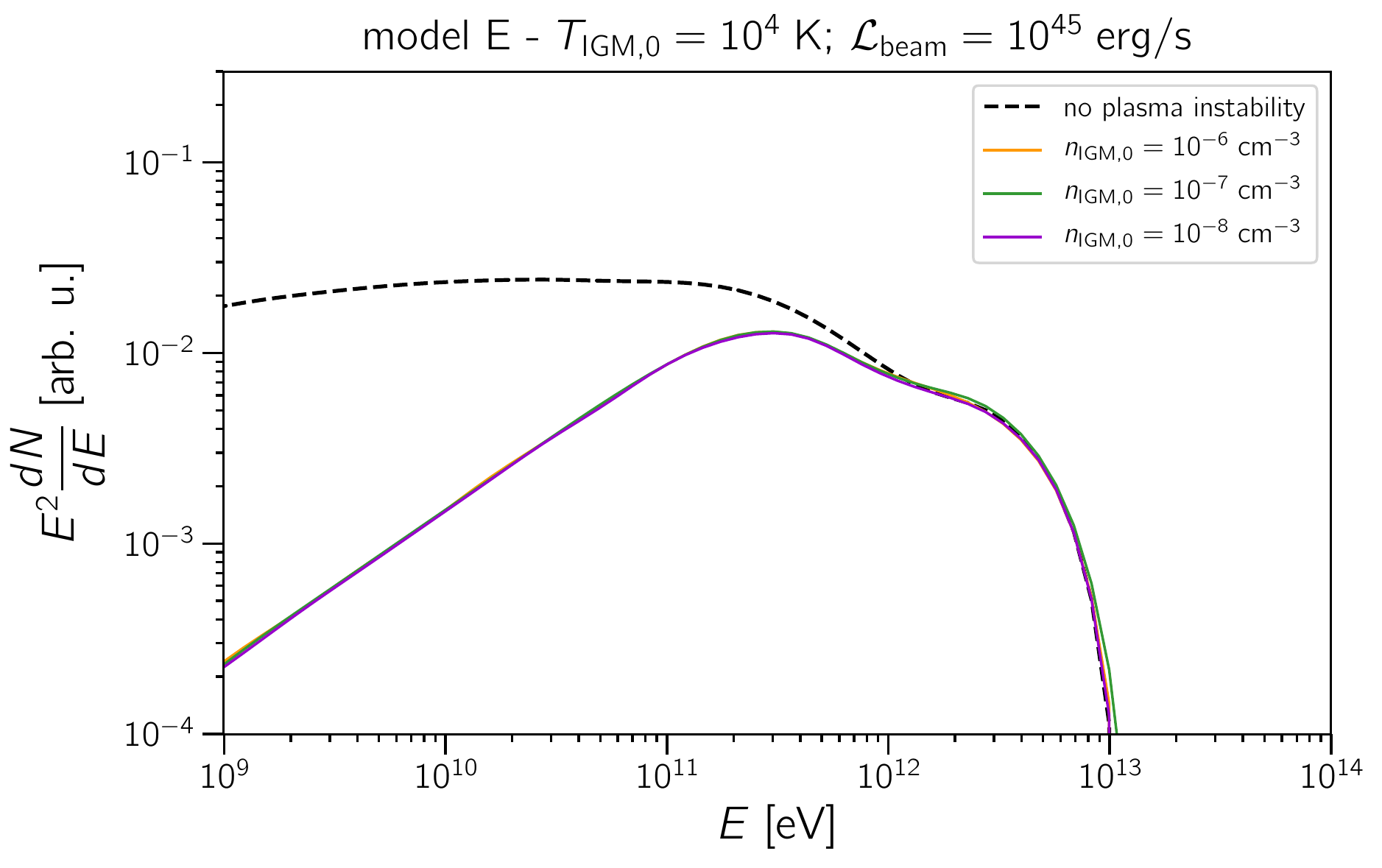}
  \caption{Gamma-ray fluxes observed at Earth for the benchmark scenario, assuming $T_{\text{IGM},0}=10^{4} \; \text{K}$ and $\mathcal{L}=10^{45} \; \text{erg/s}$. }
  \label{fig:density}
\end{figure*}

Now we discuss how the density of the IGM affects the gamma-ray spectrum. We fix $T_{\text{IGM},0}=10^{4} \; \text{K}$ and $\mathcal{L}=10^{45} \; \text{erg/s}$, varying the density only. The predicted gamma-ray spectra for our benchmark object are shown in figure~\ref{fig:density}.

Models A, D, and E are unremarkable, in the sense that the cascade is completely suppressed by the plasma instabilities, as shown in figure~\ref{fig:density}. Model C presents clear spectral signatures of the IGM density. There is a bump at $E \sim 5 \; \text{GeV}$ for $n_{\text{IGM},0}=10^{-7} \; \text{cm}^{-3}$. The onset of this feature seems to start at higher energies as  $n_{\text{IGM},0}$ increases. With our simulations we cannot confirm if this bump is shifted to lower energies, as our minimum energy\footnote{The lower energy cutoff of 1 GeV was chosen to optimise the time required to run the simulations, while  covering a range of energies that would allow for meaningful conclusions to be drawn.} is set to 1~GeV. However, if we extrapolate the analysis of the scenario with $n_{\text{IGM},0}=10^{-6} \; \text{cm}^{-3}$, it is not unreasonable to expect this to be the case. 

\subsection{Beam luminosity}

\begin{figure*}
  \centering
  \includegraphics[width=\twofigsize]{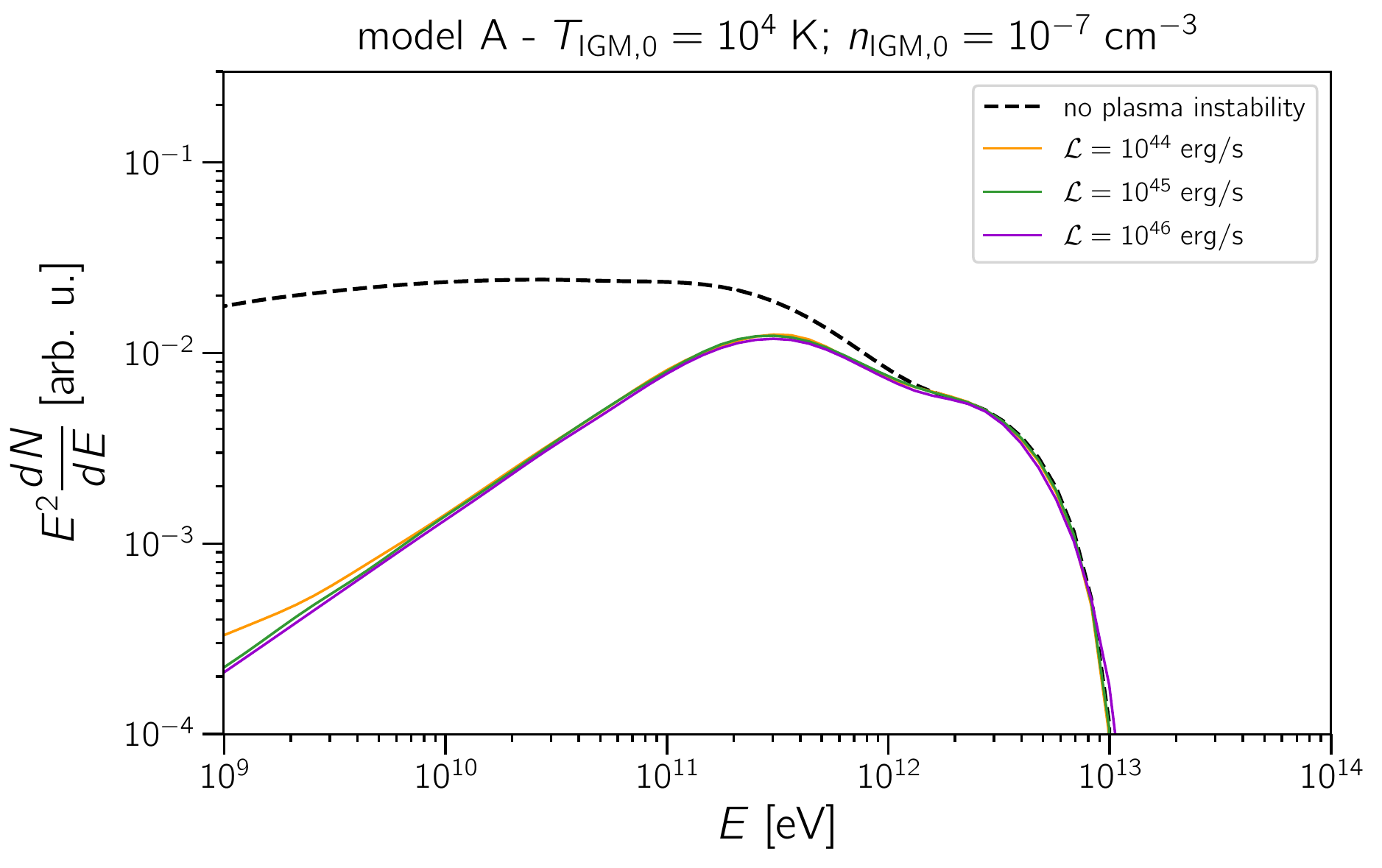}
  \includegraphics[width=\twofigsize]{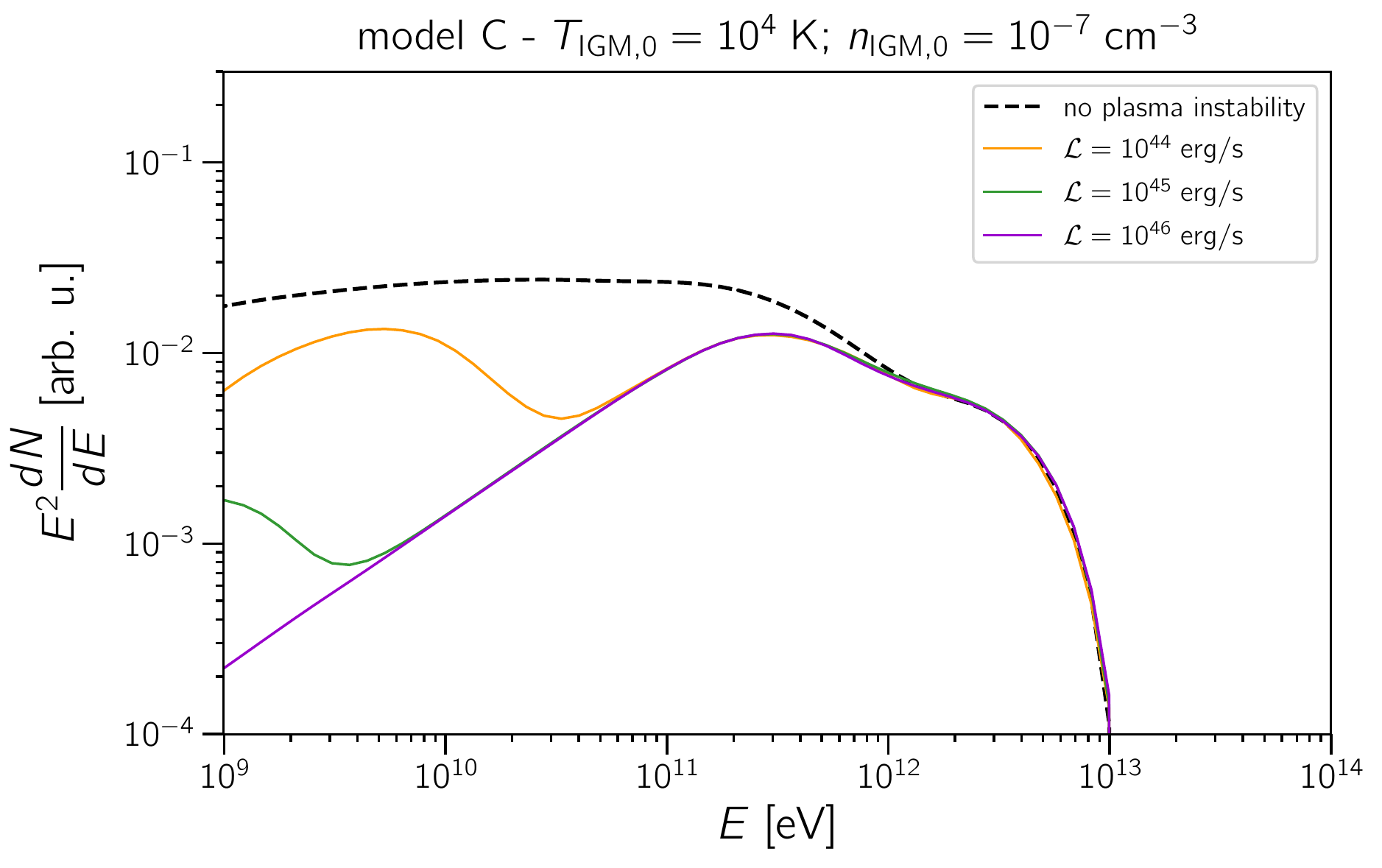}
  \includegraphics[width=\twofigsize]{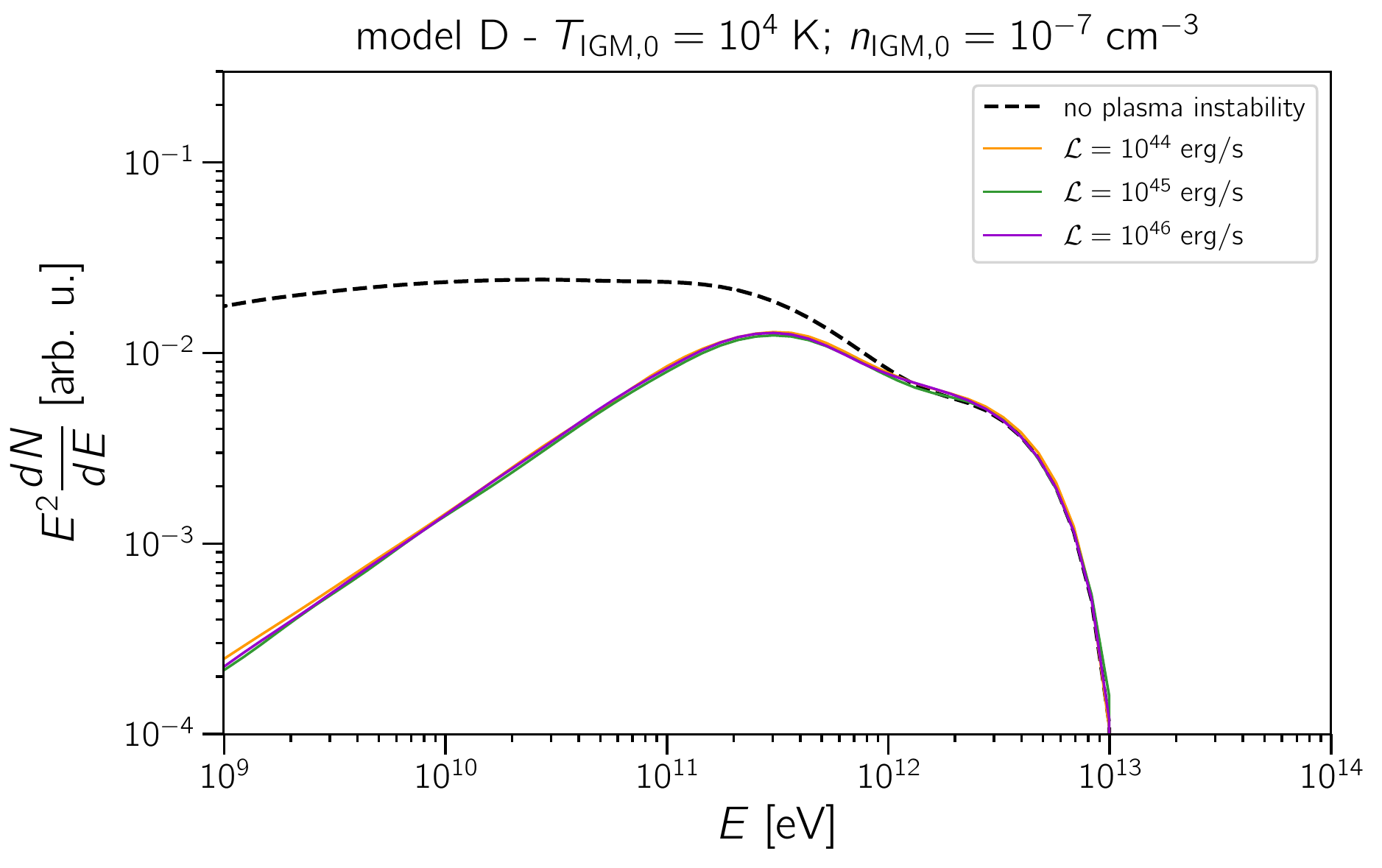}
  \includegraphics[width=\twofigsize]{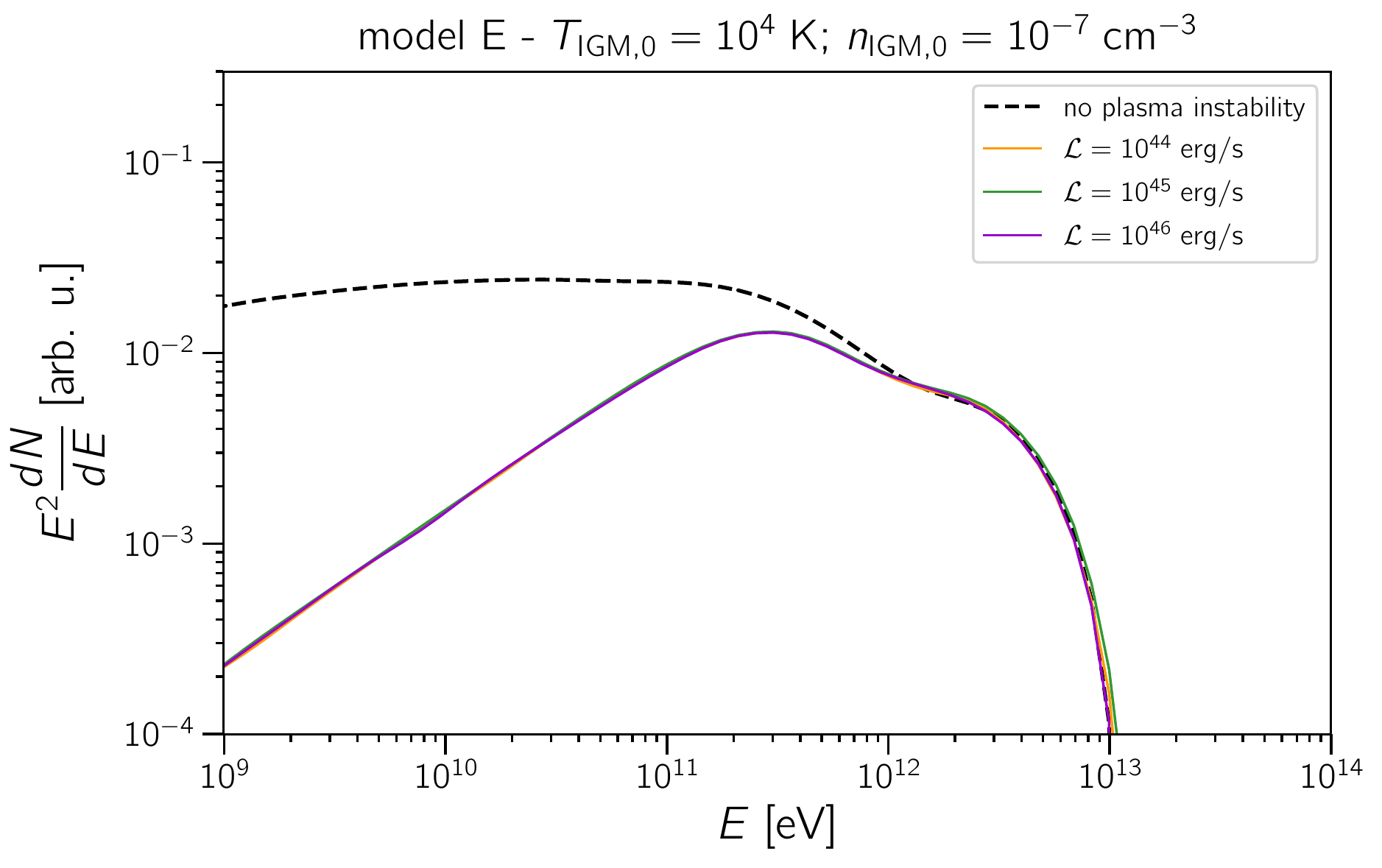}
  \caption{Gamma-ray fluxes observed at Earth for the benchmark scenario, assuming $n_{\text{IGM},0}=10^{-7} \; \text{cm}^{-3}$ and $T_{\text{IGM},0}=10^{4} \; \text{K}$. }
  \label{fig:luminosity}
\end{figure*}

We proceed our study of the relevant parameters by investigating how the luminosity of the blazar beam affects the shape of the gamma-ray spectrum. Once again, we fix all other variables and vary solely the luminosity for $T_{\text{IGM},0}=10^{4} \; \text{K}$ and  $n_{\text{IGM},0}=10^{-7} \; \text{cm}^{-3}$. The results are shown in figure~\ref{fig:luminosity}.

As in the case of the temperature and the IGM density, spectral features arise for model C and, to a lesser degree, for models A and D. As the beam luminosity decreases, an increase in the flux at $E \sim 3 \; \text{GeV}$ can be seen. The onset of this flux enhancement with respect to the full suppression of the cascade becomes visible at increasingly higher energies as the beam becomes dimmer.

The aforementioned spectral signatures are not surprising. In fact, equations~\ref{eq:tau_A}, \ref{eq:tau_C}, \ref{eq:tau_D}, and \ref{eq:tau_E} all depend on the beam density, as can be seen in equation~\ref{eq:n_beam}. 

The vast majority of blazars have isotropic-equivalent luminosities $\mathcal{L} \lesssim 10^{45.5} \; \text{erg/s}$. Therefore, the cooling due to plasma instabilities may not be as severe as for the case of very luminous objects.

\subsection{Spectral parameters}

We now investigate the effects of the spectral parameters ($E_\text{max}$ and $\alpha$) on the shape of the predicted gamma-ray spectra. To this end, we use our benchmark scenario fixing the following parameters: $T_{\text{IGM},0}=10^{4} \; \text{K}$,  $n_{\text{IGM},0}=10^{-7} \; \text{cm}^{-3}$, and $\mathcal{L}=10^{45} \; \text{erg/s}$.

As previously mentioned, for model B there are no significant spectral changes. For the other models the spectral modifications are similar to each other, so that we choose only one model to illustrate this discussion: model A.

\begin{figure*}
  \centering
  \includegraphics[width=\twofigsize]{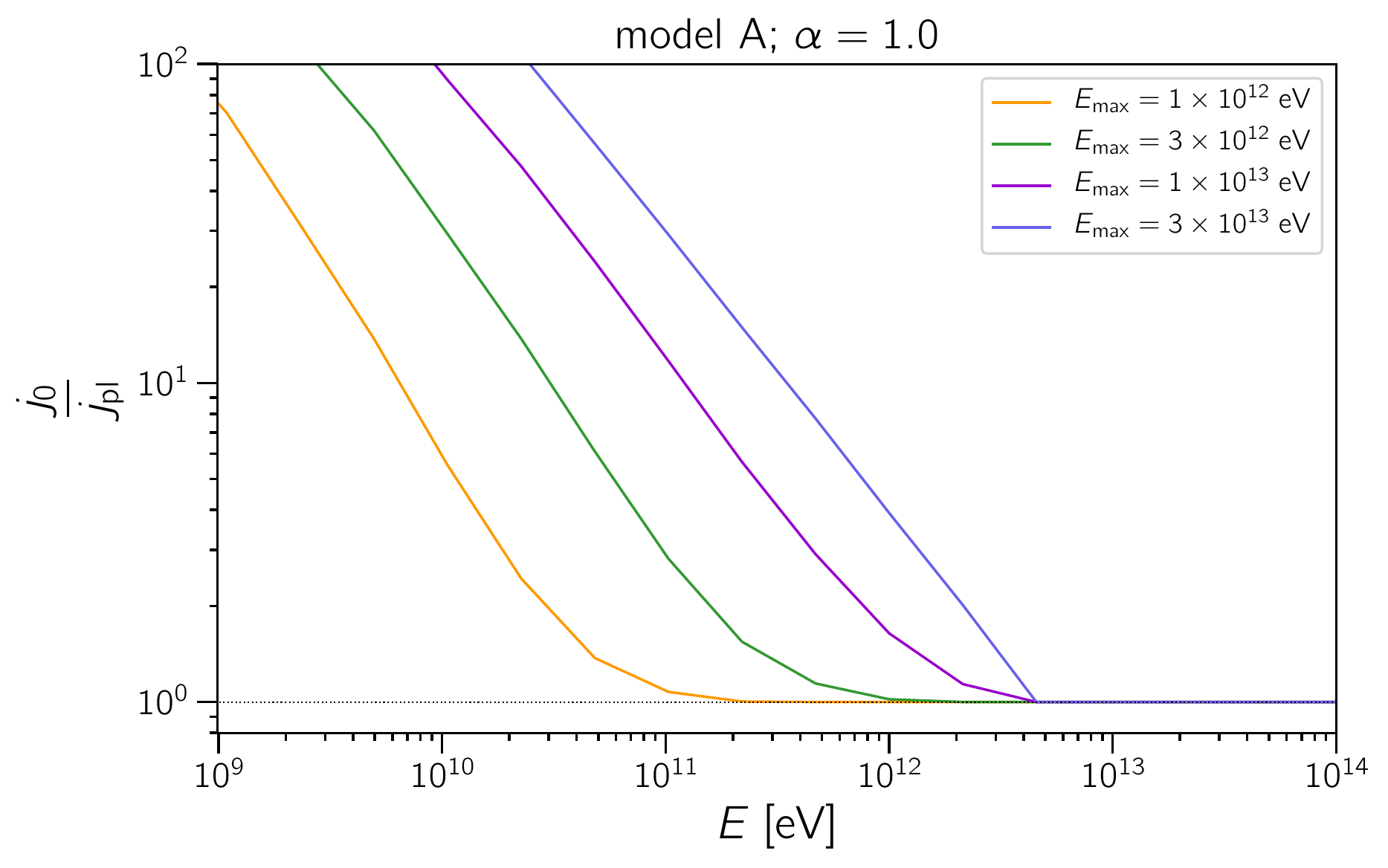}
  \includegraphics[width=\twofigsize]{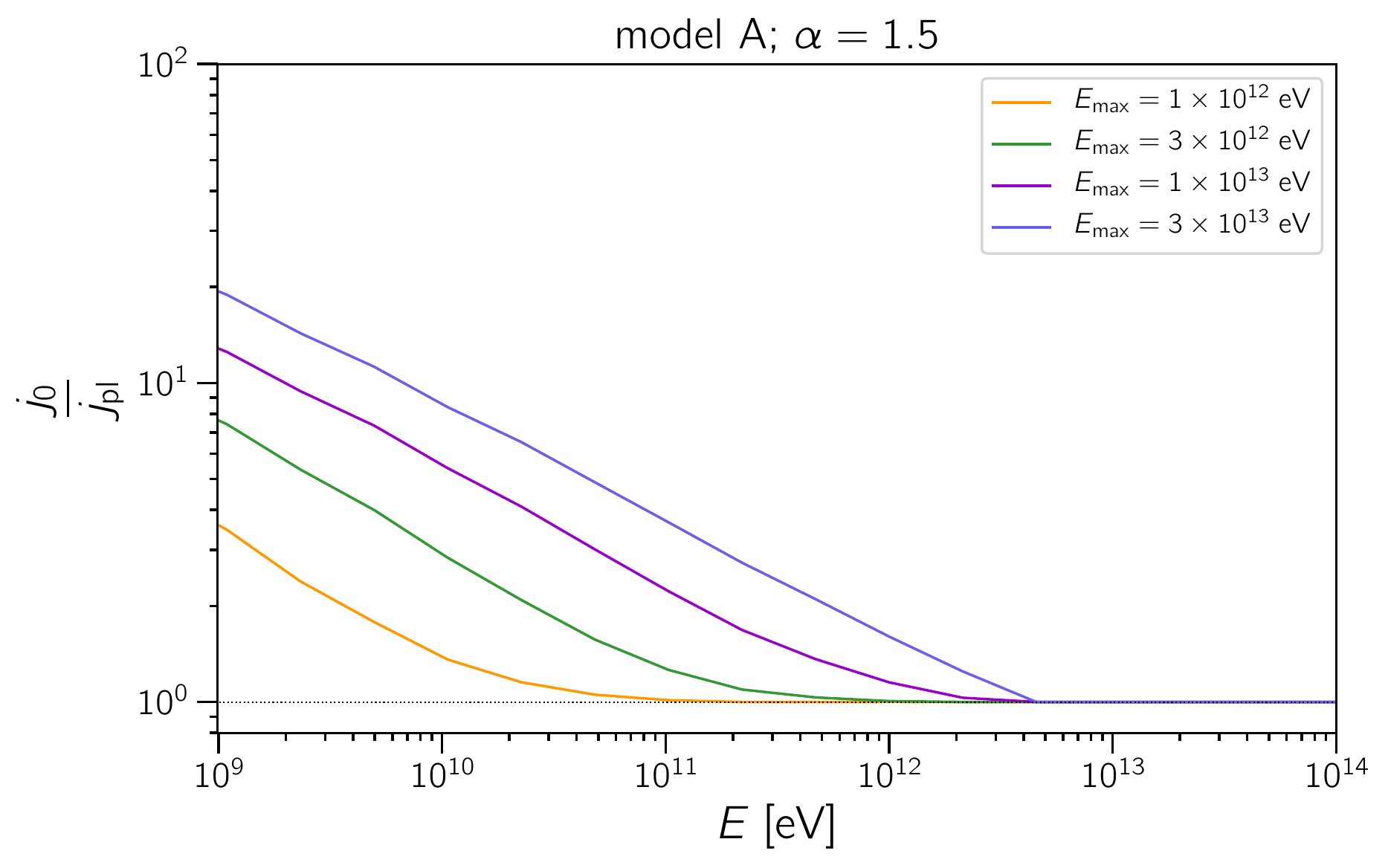}
  \includegraphics[width=\twofigsize]{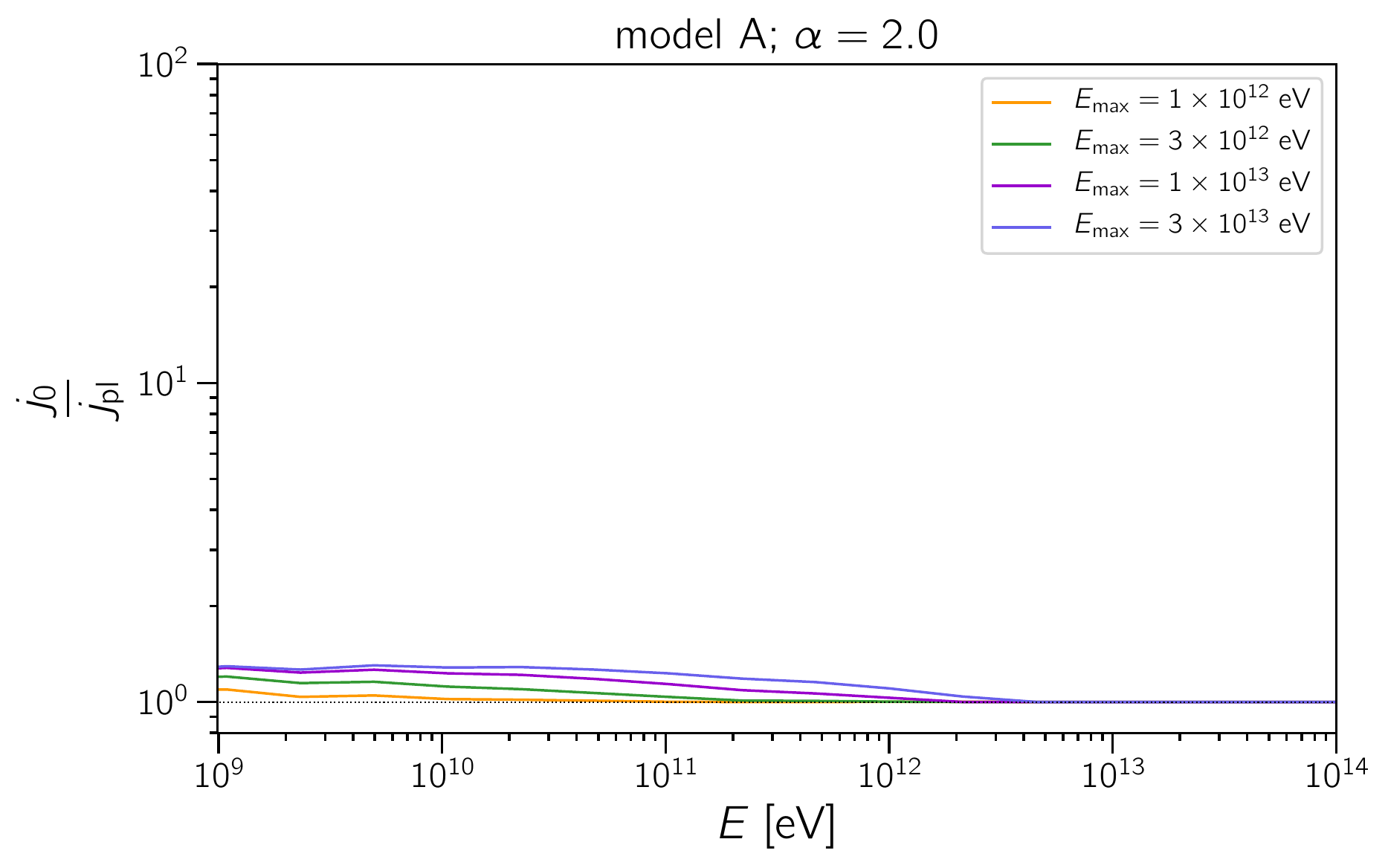}
  \includegraphics[width=\twofigsize]{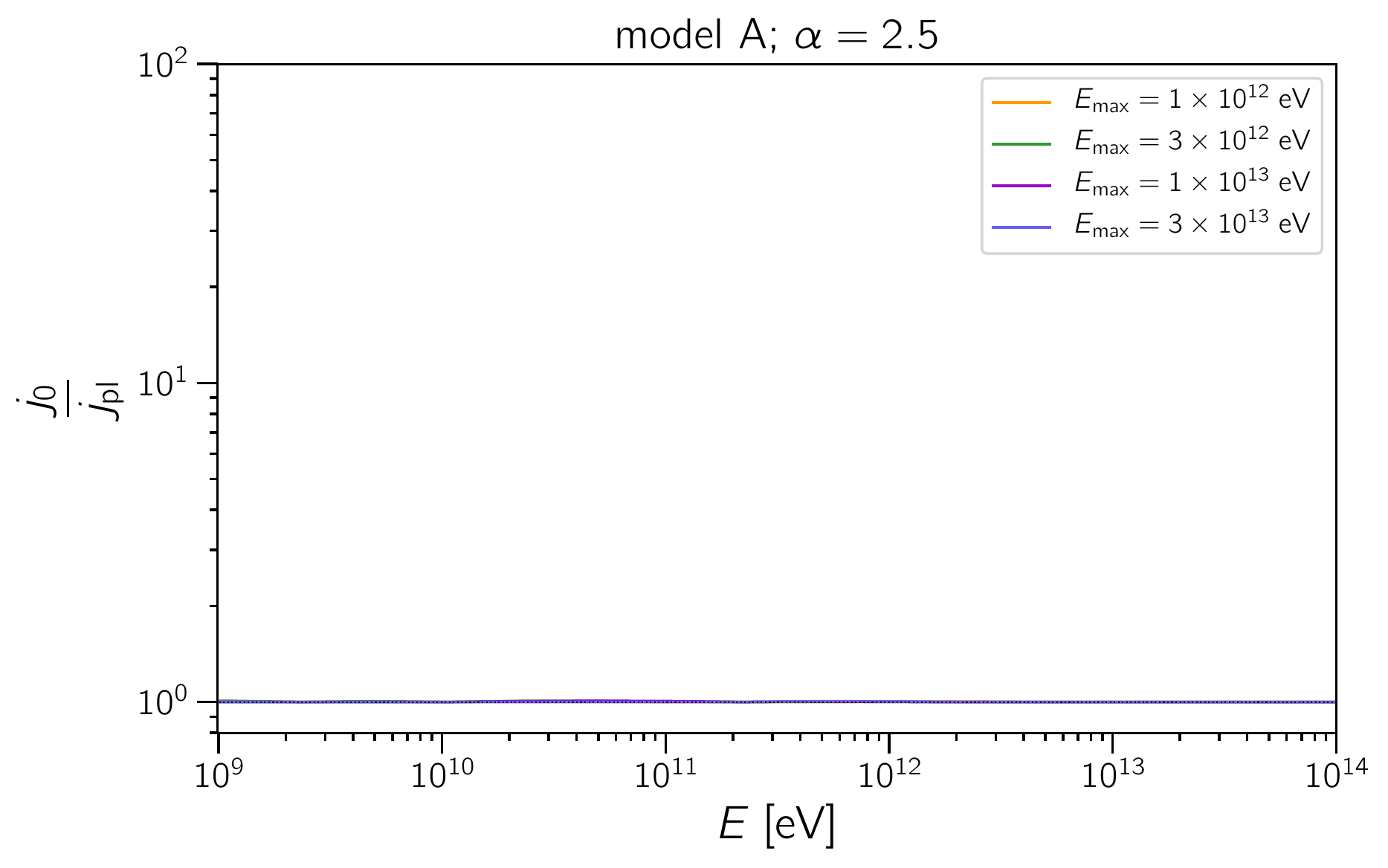}
  \caption{Gamma-ray fluxes observed at Earth for the benchmark scenario, assuming $n_{\text{IGM},0}=10^{-7} \; \text{cm}^{-3}$,  $T_{\text{IGM},0}=10^{4} \; \text{K}$, and $\mathcal{L}=10^{45} \; \text{erg/s}$. The values of the spectral indices assumed are shown at the top of each panel. A black dotted line is shown at $j_\text{pl} / j_\text{0} = 1$, for reference.}
  \label{fig:specPar}
\end{figure*}

The energy-dependent quenching factor is defined as the ratio between $j_\text{pl}$, the spectrum obtained in the presence of plasma instabilities, and $j_{0}$, the corresponding spectra in their absence, where $j \equiv dN/dE$. This is plotted as a function of the energy in figure~\ref{fig:specPar}, for different spectral indices.

We find that the blazar spectral index plays a major role in the development of the cascades. Harder spectra ($\alpha \lesssim 1.5$) lead to quenching factors of $\sim 100$ at $E \sim 1 \; \text{GeV}$ for $E_\text{max} \lesssim 10 \; \text{TeV}$, whereas for $\alpha \gtrsim 2$ the quenching factors are $\lesssim 2$. The maximal energy also significantly affects the spectral shape. The higher the $E_\text{max}$, the larger the quenching factors for a fixed spectral index.

These results are expected. The contribution of events with higher energies is larger in the case of harder spectra (lower $\alpha$) than for softer spectra. Similarly, as $E_\text{max}$ increases, so does the quenching factor.

\section{Application to selected blazars}\label{sec:blazars}

In this section we discuss the consequences of plasma instabilities for VHEGR spectra of a few blazars. We select a few extreme BL Lacertae objects known to produce multi-TeV gamma rays. This list is not meant to be complete, being merely a sample of objects commonly used in studies aiming to constrain IGMFs with gamma rays. We list these object in table~\ref{tab:objects}.

\begin{table*}
\centering
\caption{Objects used in this study, together with values for the parameters of their intrinsic spectrum that are compatible with observations, as taken from the corresponding references.}
\begin{tabular}{cccccc}
\hline
\hline
object & z & $\alpha$ & $E_\text{max}$ [TeV] & observations & parameters \\
\hline
1ES 0229+200  & 0.140 & 1.2 & 5.0 & \citet{Taylor:2011bn} & \citet{Taylor:2011bn} \\
1ES 0347-121  & 0.188 & 1.8 & 3.0 & \citet{2015MNRAS.451..611B} & \citet{2018MNRAS.477.4257C} \\
1ES 0414+009  & 0.287 & 1.9 & 2.0 & \citet{2012ApJ...755..118A} & \citet{2018MNRAS.477.4257C} \\
1ES 1101-232  & 0.186 & 1.7 & 4.0 & \citet{Abramowski:2014uta} & \citet{2018MNRAS.477.4257C} \\
1ES 1218+304  & 0.182 & 1.9 & 2.0 & \citet{Taylor:2011bn} & \citet{2018MNRAS.477.4257C} \\
1ES 1312-423  & 0.105 & 1.9 & 1.3 & \citet{2013MNRAS.434.1889H} & \citet{2013MNRAS.434.1889H} \\
RGB J0710+591 & 0.125 & 1.9 & 2.0 & \citet{Taylor:2011bn} & \citet{2018MNRAS.477.4257C} \\
PKS 2155-304  & 0.116 & 1.9 & 10.0 & \citet{Abramowski:2014uta} & \citet{Abramowski:2014uta} \\
\hline
\end{tabular}
\label{tab:objects}
\end{table*}

Note that the parameters $\alpha$ and $E_\text{max}$, presented in table~\ref{tab:objects}, are used only to illustrate the effects of the plasma instabilities on the gamma-ray flux and for a qualitative discussion. It is beyond the scope of this work to perform a fit of the model to the data.

\begin{figure*}
    \centering
    \includegraphics[width=\twofigsize]{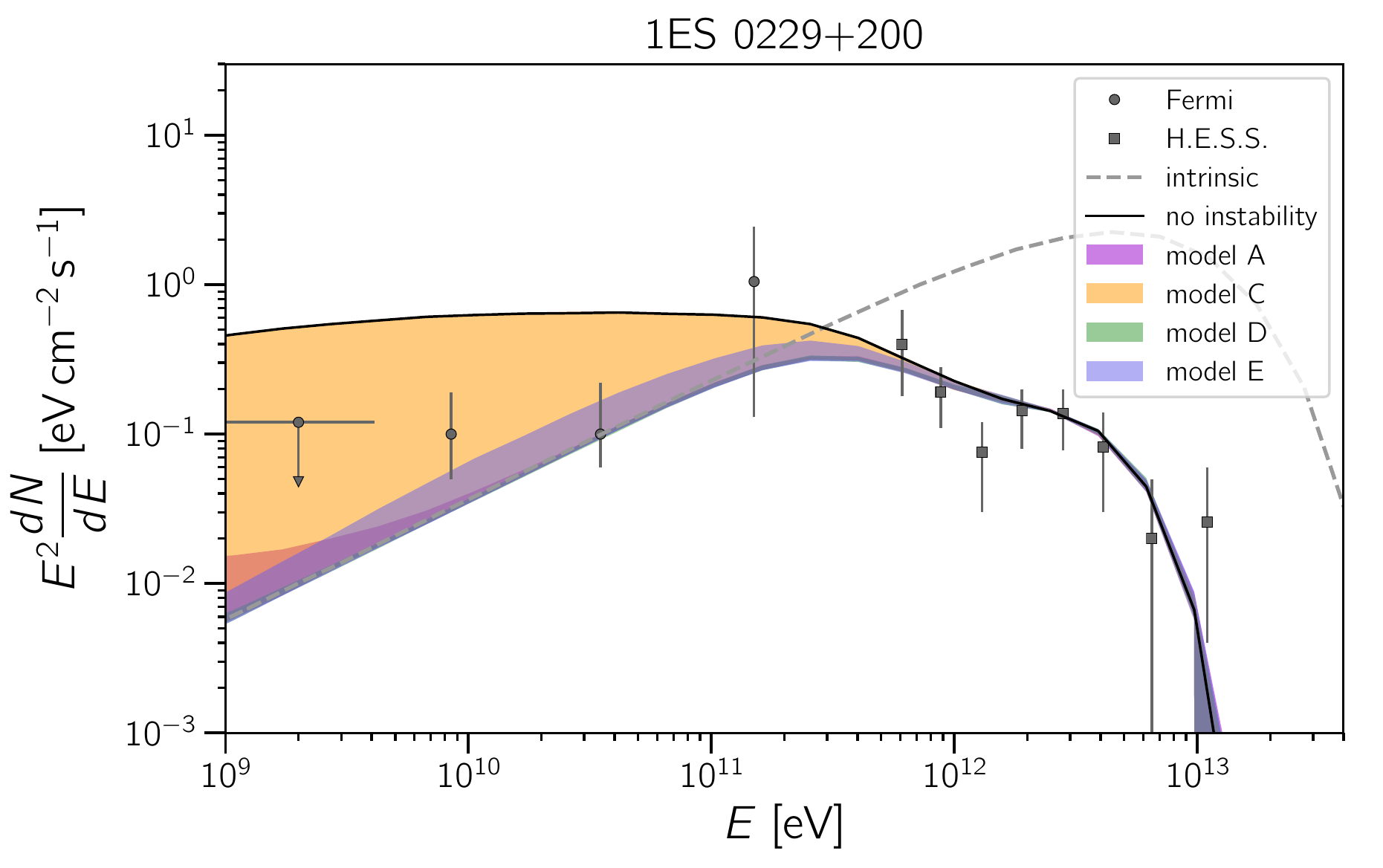}
    \includegraphics[width=\twofigsize]{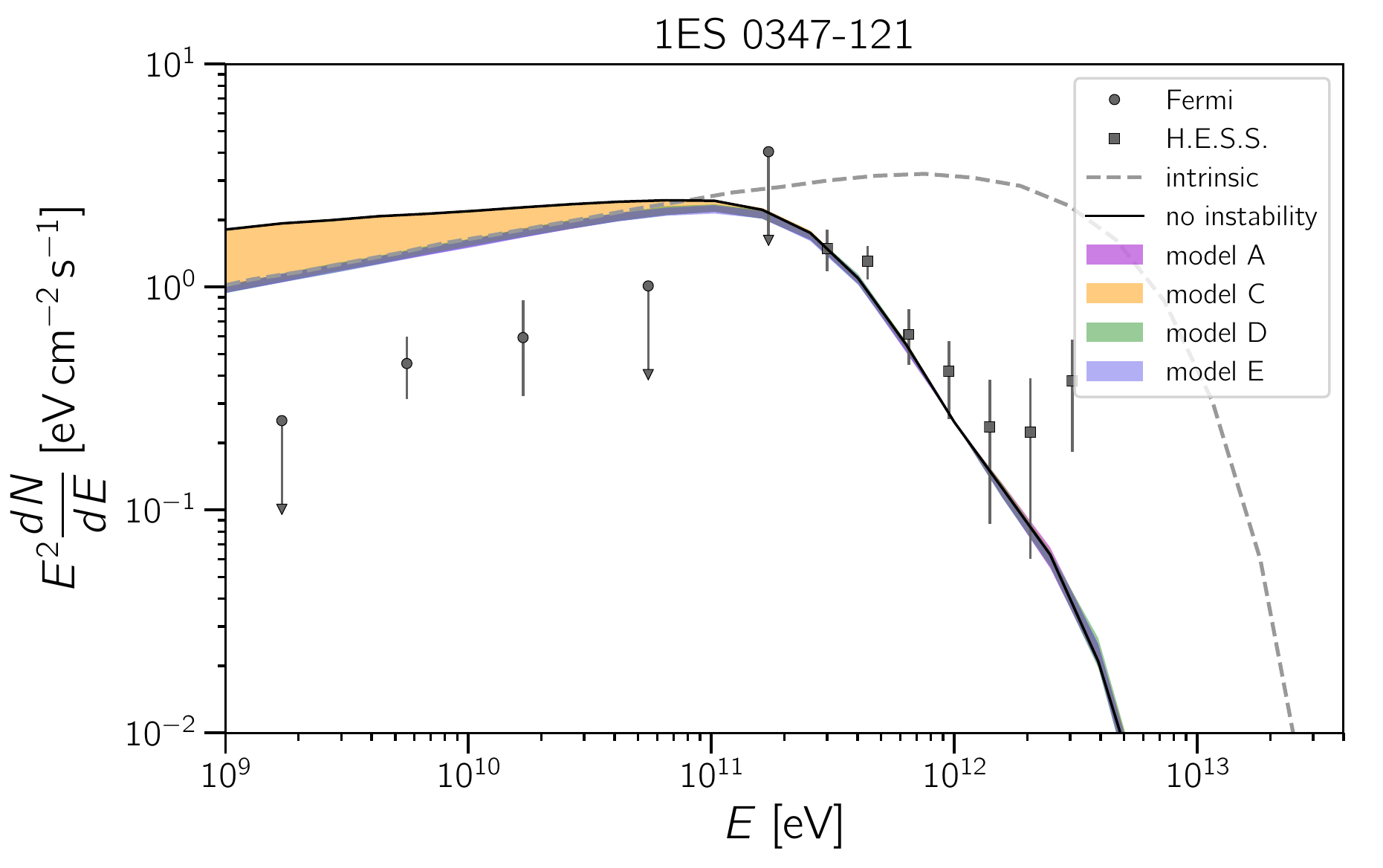}
    \includegraphics[width=\twofigsize]{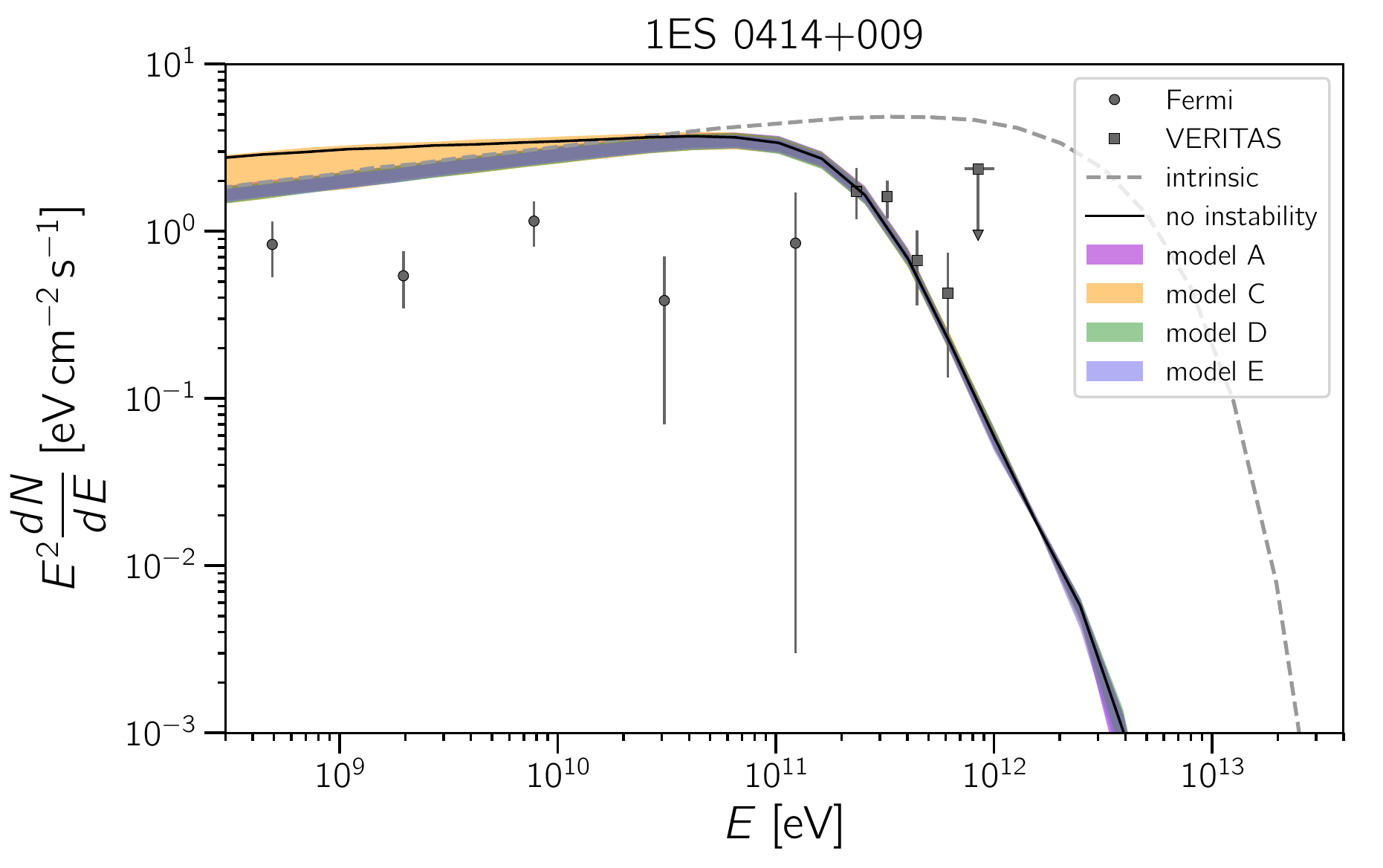}
    \includegraphics[width=\twofigsize]{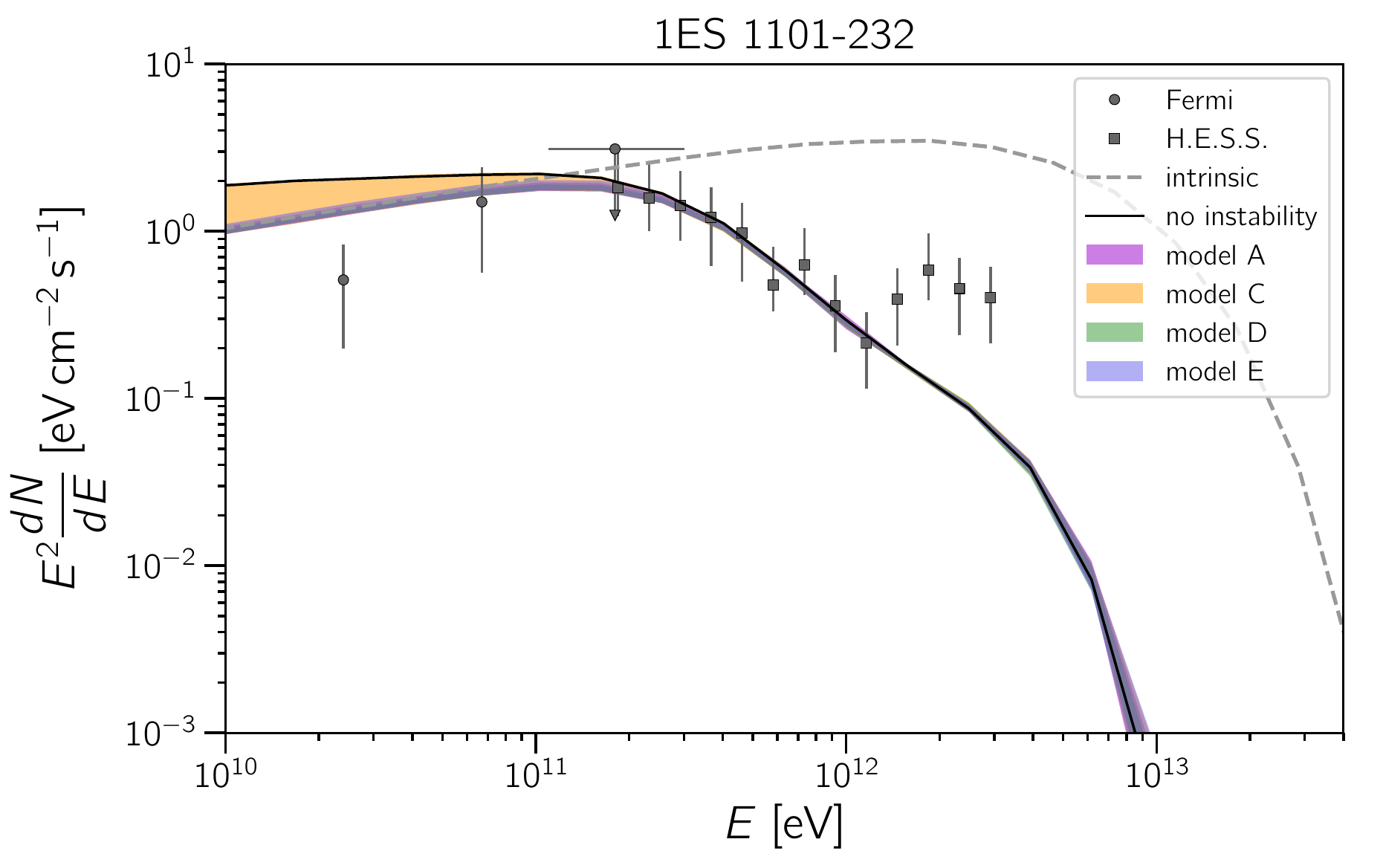}
    \includegraphics[width=\twofigsize]{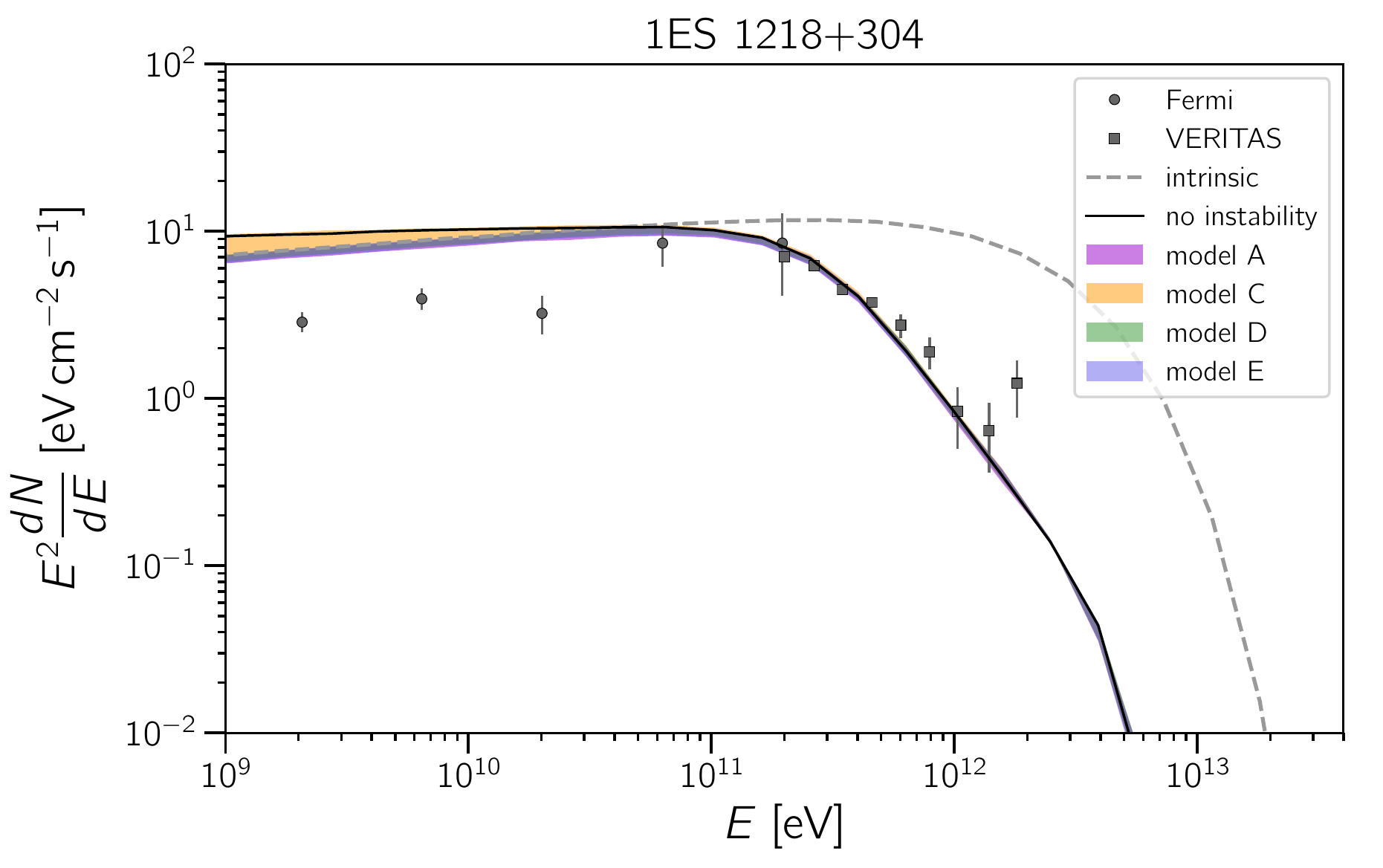}
    \includegraphics[width=\twofigsize]{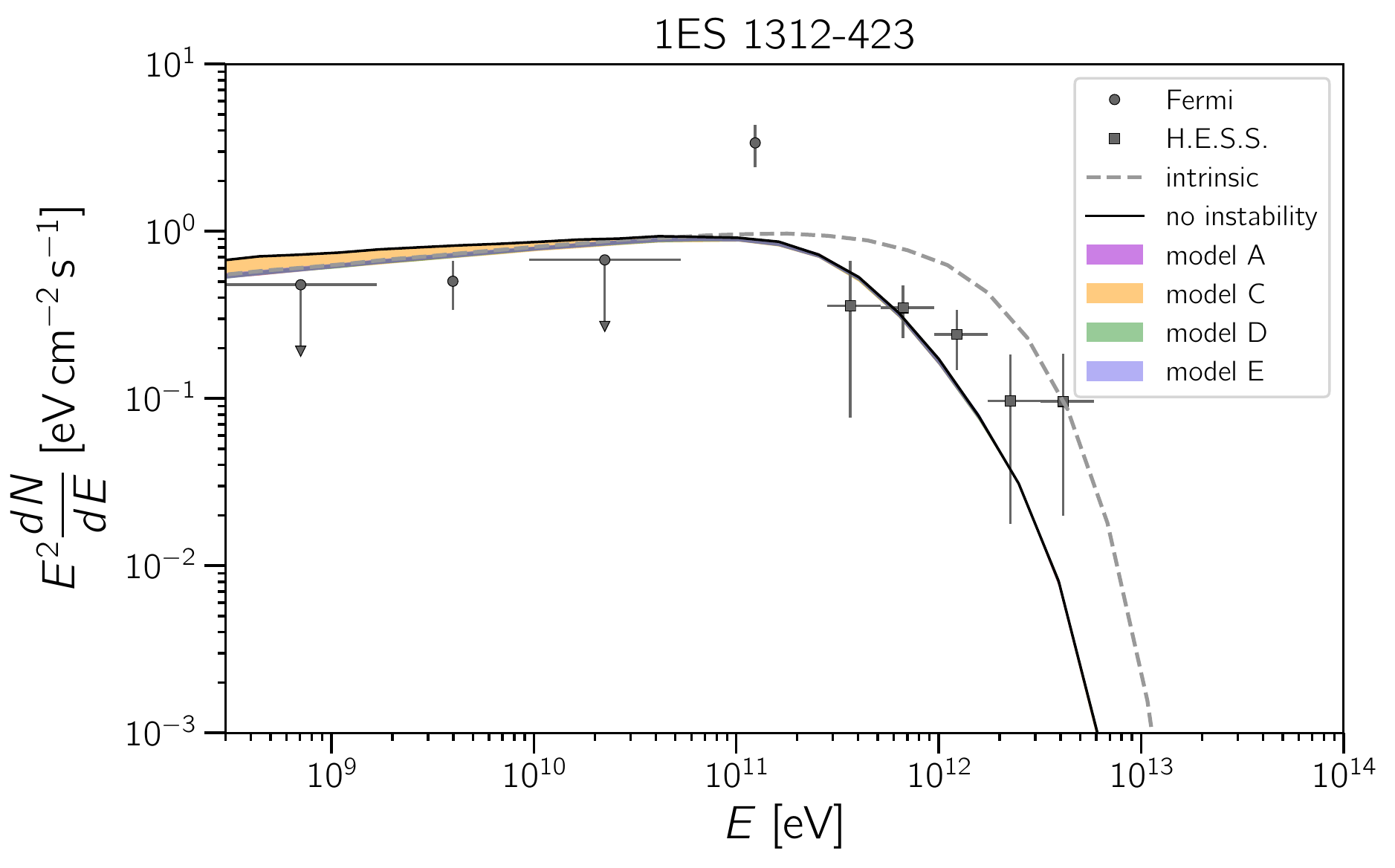}
    \includegraphics[width=\twofigsize]{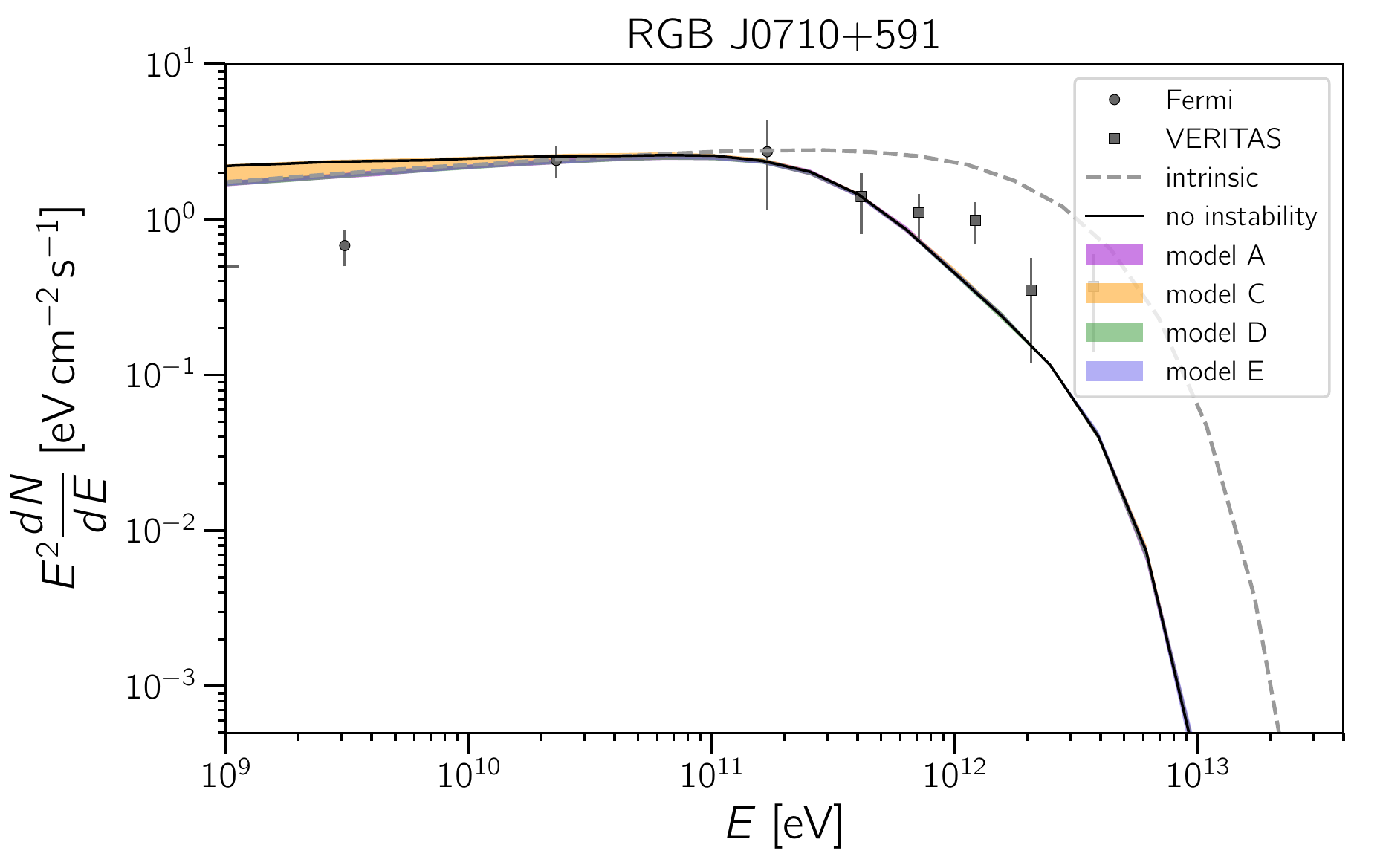}
    \includegraphics[width=\twofigsize]{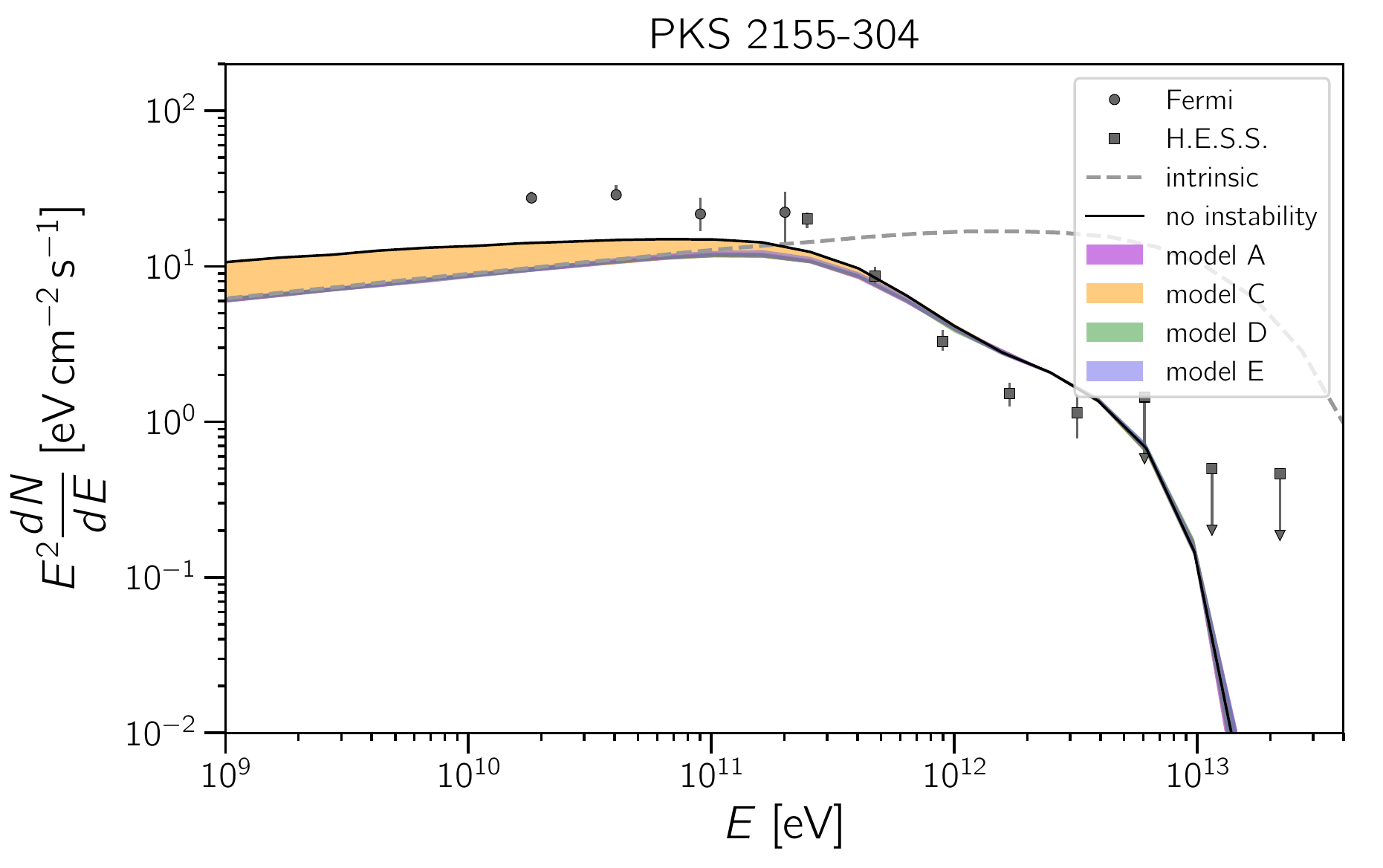}
    \caption{Simulated gamma-ray fluxes predicted for the blazars listed in table~\ref{tab:objects}. The different models of plasma instabilities including a range of uncertainties in the parameters $\mathcal{L}$, $T_\text{IGM}$, $n_\text{IGM}$ (coloured bands). The solid black line represents the case without the instabilities. The parameters corresponding to the intrinsic spectrum of the objects are also presented in table~\ref{tab:objects}, along with the references for the data points. The dashed line represents the intrinsic source spectrum.}
    \label{fig:objects}
\end{figure*}

In figure~\ref{fig:objects} we compute the expected gamma-ray fluxes considering the effect of plasma instabilities. As discussed in section~\ref{sec:parameters}, the predictions for model B are compatible with a negligible action of the instabilities and are ignored in this section. We consider the range of parameters discussed in the section~\ref{sec:parameters} and represent these uncertainties as bands.

From figure~\ref{fig:objects} we can draw our first conclusion, namely that the cooling due to the instabilities can cause a substantial hardening of the spectrum at $E \lesssim 100 \; \text{GeV}$. The spectral signatures of this effect are similar to those of IGMFs.

We note that the spectral suppression due to the plasma instability depends on the intrinsic spectrum of the source. For instance, this effect is small for objects like 1ES 1218+304 and 1ES 1312-423, whose (fitted) spectral indices are $\alpha \approx 1.9$. On the other hand, for 1ES 0229+200, whose spectral index is $\alpha \approx 1.2$~\citep{Taylor:2011bn}, the predicted flux varies by several orders of magnitude at $E \lesssim 100 \; \text{GeV}$. For this reason, the spectra of most of the objects shown in figure~\ref{fig:objects} could change considerably, improving the agreement with the observations.

For 1ES 0229+200, \citet{2018MNRAS.477.4257C} found that $\alpha = 1.5$ and $E_\text{max}=12 \; \text{TeV}$ adjusts the data satisfactorily. On the one hand, the increase in the spectral index would decrease the width of the bands and mitigate the effects of plasma instabilities; on the other hand, the increase in the maximal energy from 5 TeV to 12 TeV  may increase this effect as $\alpha$ and $E_\text{max}$ are degenerate parameters.

We have considered blazars whose variabilities are small over the period of time of observation, with exception of PKS 2155-304. This object presents a state of enhanced emission during a flaring episode whose contribution is excluded from the analysis, as done by~\citet{Abramowski:2014uta}. 

The uncertainty due to the choice of the EBL model (\citet{Gilmore:2011ks} in this work) does affect our results quantitatively. Qualitatively, the same conclusions as the ones presented here hold. In this case, the breadth of the band could be enlarged or reduced.

The luminosity of the beam emitted by the object can affect the observed gamma-ray flux, as shown in figure~\ref{fig:luminosity}. This is particularly important for model C. Nevertheless, all objects listed here have very similar luminosities, of the order of $\mathcal{L} \sim 10^{43.5} - 10^{45} \; \text{erg/s}$ (see e.g.~\cite{0004-637X-752-1-22}).

One could think that for the combinations of $\alpha$ and $E_\text{max}$ adopted, it would be possible to constrain some plasma instability models using the plots from figure~\ref{fig:objects}. However, our calculations are meant to maximise the effects of the instabilities. A change in their growth time, for instance, could render the plasma cooling much more inefficient.

\section{Discussion}\label{sec:discussion}

The relevance of plasma instabilities for the development of electromagnetic cascades is a rather controversial issue. Many authors have constrained IGMFs neglecting the possible existence of the instabilities.  

In their seminal work \citet{0004-637X-752-1-22} claimed that the effects of plasma instabilities place stringent bounds on the strength of IGMFs, $B \lesssim 10^{-12} \; \text{G}$. This limit was derived by comparing the growth time of the oblique instability with the Larmor radius described by the electrons. This argument holds approximately, and other factors such as the magnetic power spectrum may play an important role in determining the propagation regime of the electrons. For instance, if the coherence length of the IGMF were much smaller than the cooling length of the charged component of the cascade, then plasma instabilities could not arise as the initial beam would be disrupted to the point where it ceases to be collimated, since the particles do not move in a specific direction, but rather propagate via (magnetic) diffusion.

An important issue raised by~\citet{DuNe} is that \citet{0004-637X-752-1-22}, in their original work, have limited their analysis to the the linear approximation, neglecting the backreaction of the beam perturbations on the growth rate. As a consequence, oblique modes of Langmuir waves would be suppressed due to non-linear Landau damping, stabilising the beam, and thus minimising the role plasma instabilities would play on the development of the cascade. This argument explains why plasma effects on the gamma-ray spectrum are not visible for model B; they are in line with the findings of \citet{Miniati:2012ge}. The same arguments are made by \citet{0004-637X-758-2-102,Schlickeiser:2013eca}, though the range of parameters for which it holds is different, explaining why in this scenario (model C) the cascade is quenched by the instabilities for most combination of parameters.

\citet{Yan:2018pca} claimed that even if plasma instabilities are considered, in particular a fast-growing oblique instability, IGMFs could still be reliably constrained using electromagnetic cascades. In this case, the IGMF strength would change by a factor $\sim 10$. At first sight, these results do not seem to agree with ours. Nevertheless, one should bear in mind that our results are limiting cases aiming at maximising the quenching of the cascade by the plasma instabilities (equation~\ref{eq:taui}). Therefore, if we assume that only a small fraction of the total energy is deposited in the IGM, i.e., if the instability is not as efficient as we considered, than our results do not conflict with those by \citet{Yan:2018pca}. More studies using three-dimensional simulations are required to draw further conclusions.

The modelling of the gamma-ray flux considering plasma instabilities is non-trivial in the presence of IGMFs. For instance, the Weibel instability (subdominant and not explicitly treated here) is known to generate and amplify magnetic fields, as observed in laboratory  experiments~\citep{2015NatPh..11..173H}. This may disrupt the beam, increasing its transverse momentum and potentially suppressing the instability. In fact, the momentum distribution of the pairs seems to be crucial to estimate the evolution of the instabilities. 

Care should be taken when stating whether or not plasma instabilities quench the development of electromagnetic cascades in the IGM. Multiple kinds of instabilities can arise, and under certain conditions their cooling rates may be subdominant with respect to inverse Compton scattering. This, for instance, is the case for model C if the luminosity of the blazar beam is $\lesssim 10^{44} \; \text{erg/s}$, or if the electron-positron pairs are created in a relatively cold region of the IGM ($T_\text{IGM} \sim 10^3 \; \text{K}$), or if the density of the IGM is exceedingly high. Similarly, some dependence on the temperature can also be seen for model E, although in none of the cases studied the effects of the instabilities are negligible.

Caveats exist in our analysis. First, we consider that the cooling is completely effective, which may not be true if the instability grows slower than predicted. In this sense, our results should be taken as a limiting case wherein energy dissipation via plasma instabilities is fully efficient. Longer growth times would decrease the energy-loss rates assumed, modifying the cooling rates shown in figure~\ref{fig:rate} such that inverse Compton scattering could become the dominant cooling process. Our model is one-dimensional. The effects of IGMFs were  purposely neglected in the this study in order to maximise the effects of the plasma instabilities.  In the presence of IGMFs, the distribution of momentum of the electrons would suffer some change. This could, in principle, reduce the role played by the instabilities.

In our analysis only the EBL model by \citet{Gilmore:2011ks} was considered. Although qualitatively the discussion would remain unaltered if we adopted another EBL model, quantitavely it would change. In particular, the contribution of events with $E \gtrsim E_\text{max}$ would be more pronounced for EBL models wherein the density of target infrared photons is lower. Ultimately, it all depends on an interplay between the injected spectrum ($E_\text{max}$ and $\alpha$), and the distance of the object, which follows immediately from figure~\ref{fig:ratePP}. As for the~GeV part of the observed spectrum, if the flux is completely dominated by cascade photons, we expect it to retain the slope corresponding to the spectral index of the source, regardless of the choice of EBL model.

Our results suggest that the quenching factors are strongly dependent on the spectral index and maximal energy of the blazar. These parameters are determined by the underlying acceleration mechanism in the source, as well as by its opacity. Extreme blazars are thought to have spectra $\alpha \gtrsim 1.5$~\citep{hess2006a,cerruti2015a,2015MNRAS.451..611B}. Our results from figure~\ref{fig:objects} suggest that the effects of plasma instabilities are relatively small for our choices of $\alpha$ and $E_\text{max}$, with the exception of the object 1ES 0229+200, whose $\alpha$ is the smallest among the sources studied ($\alpha = 1.2$). Therefore, since $1.5 \lesssim \alpha \lesssim 2.0$ for most extreme blazars, the suppression factor may be lower than $\sim 10$; for $\alpha \gtrsim 1.8$ the quenching factor is relatively small and we conclude that, unless $E_\text{max}$ is high, then the spectral changes due to the instabilities are small. 

Taking the parameters from table~\ref{tab:objects}, we see that plasma instabilities do not explain the spectra of some blazars such as 1ES~0347-121, 1ES~0414+009, 1ES~1101-232, and 1ES~1218+304, as can be seen in figure~\ref{fig:objects}. This leaves room for IC cooling of the pairs to dominate, and for IGMF signatures to be seen in blazar spectra. A detailed fit of the fluxes of these objects measured by the Fermi telescope and IACTs for different combinations of $n_\text{IGM}$ and $T_\text{IGM}$ could change the values of the best-fit of $\alpha$ and $E_\text{max}$; it is not within the scope of this work to perform such an analysis. Nevertheless, it is unlikely that the fitted maximal energy would change considerably if the model is altered to include plasma instabilities.
As a consequence, as one can see in figure~\ref{fig:specPar}, the quenching factors would not be enough to explain the fluxes from the aforementioned objects for this set of parameters.

We have treated $n_\text{IGM}$ and $T_\text{IGM}$ as independent quantities. In reality, their evolutions are correlated as the temperature affects the reionisation of the gas and consequently the density of the IGM plasma~\citep{2001ApJ...562...52M}. This relation may be useful to decrease the number of degrees of freedom when attempting to probe the IGM with VHEGRs.  

The energy dissipated by plasma instabilities may be absorbed by either the IGM, thus causing beam cooling, or by the beam itself, leading to its disruption~\citep{0004-637X-752-1-22}. This difference is not important when it comes to interpreting the spectra of TeV-emitting blazars, but it is of fundamental importance for discussing other observational consequences of this phenomenon such as its effect on structure formation~\citep{2012ApJ...752...24P} and on IGM heating~\citep{Chang:2011bf}. In particular, the thermal history of the Universe depends on the fraction of the total energy that is used to heat the IGM; this efficiency factor is $\sim 10\%$ for model~D, and $\sim 50$--$100\%$ assuming model~A.

Our analysis focuses solely on blazars. In the context of the AGN paradigm, blazars have their jets pointing directly at Earth. Small deviations from the line of sight are not unexpected. Moreover, the jets have characteristic opening angles of $\sim 5^\circ$. This misalignment of the jet emission, too, may affect the growth of instabilities, since the density of the beam decreases as the distance to the axis of the jet increases. While this effect is likely small, a careful modelling of the cascade and three-dimensional simulations would be required to confirm this picture for the specific case of misaligned blazars.  

For typical blazars, only after $\sim 300$ years~\citep{0004-637X-752-1-22} can the instabilities grow enough to be able to cool the electron-positron pairs. Thus, flaring objects may not exhibit the same kind of behaviour as steady sources. As a consequence, they may still be reliable probes of IGMFs. 

The spectral changes stemming from plasma instabilities or IGMFs are not unique. \citet{Essey:2009ju,Essey:2010nd,Essey:2010er} have suggested that ultra-high-energy cosmic rays (UHECRs) can induce electromagnetic cascades in the IGM, with similar observational signatures. 

\section{Conclusions and Outlook}\label{sec:conclusions}

We developed a numerical tool to estimate the cooling rate of electrons due to plasma instabilities caused by interactions between a beam of very energetic particles emitted by blazars. We compared different kinds of instabilities and various models found in the literature. We applied this tool to simulate the spectrum of several blazars.

Our main conclusion is that the blazar spectrum determines the impact of plasma instabilities on the development of electromagnetic cascades in the IGM. In other words, the relevance of the instabilities is related to an interplay between the blazar spectral index and maximal energy. Therefore, a detailed understanding of the VHE emission by blazars is required to determine whether plasma instability dominates over inverse Compton cooling.

We stress here, once more, that our results have to be understood as the limiting case under the assumption of maximal growth of the dominant instability for each model, as our goal was to derive robust lower limits for the observed photon flux. In the future, we intend to simulate more realistic cases and account for other processes which may decrease the influence of the instabilities on the development of electromagnetic cascades.

We have shown that for typical IGM parameters and beam luminosities some types of plasma instabilities may cool electron-positron pairs faster than inverse Compton scattering. As a consequence, these instabilities may lead to a hardening of the spectrum of blazars at energies $\lesssim 100 \; \text{GeV}$. This effect resembles the suppression of the flux caused by IGMFs. Therefore, IGMF constraints may be compromised due to a possible dominance of plasma instability cooling over inverse Compton scattering during the development of electromagnetic cascades in the IGM. This result, however, depends on the hardness of the intrinsic spectrum of the blazar.

The existence of pervasive IGMFs cannot be excluded. Therefore, future studies combining both the effects of IGMFs and plasma instabilities are required to predict gamma-ray fluxes from blazars more accurately, especially for those sources for which the plasma instabilities (even overestimated) do not seem to be responsible for the hardening  of the $\lesssim 100$ GeV part of the spectrum (i.e., 1ES~0347-
121, 1ES~0414+009, 1ES~1101-232, and 1ES~1218+304, as 
seen in figure \ref{fig:objects}). Moreover, efforts towards simulating the growth of these instabilities using magnetohydrodynamical or particle-in-cell simulations would be needed to unambiguously determine whether or not they can quench electromagnetic cascades.

An uncontroversial window of opportunity for constraining IGMFs with electromagnetic cascades remains. The search for echoes associated with transient events, i.e., magnetically-induced delays in the arrival secondary gamma rays from primary VHEGR emission, are still excellent probes of IGMFs because the effects of plasma instabilities may be small. This is the case for some particular energetic events in AGNs. Gamma-ray bursts seem very promising candidates for this purpose.

\section*{Acknowledgments} 

RAB is supported by grant \#2017/12828-4, S\~ao Paulo Research Foundation (FAPESP). The work of AS was supported by the Russian Science Foundation under grant no.~19-71-10018, carried out at the Immanuel Kant Baltic Federal University. EMGDP is partially supported by FAPESP grant \#2013/10559-5 and CNPq grant 306598/2009-4. The simulations were run at the cluster GAPAE of the group of Plasmas and High-Energy Astrophysics of IAG-USP (FAPESP grant \#2013/10559-5).






\begin{thebibliography}{}
\makeatletter
\relax
\def\mn@urlcharsother{\let\do\@makeother \do\$\do\&\do\#\do\^\do\_\do\%\do\~}
\def\mn@doi{\begingroup\mn@urlcharsother \@ifnextchar [ {\mn@doi@}
  {\mn@doi@[]}}
\def\mn@doi@[#1]#2{\def\@tempa{#1}\ifx\@tempa\@empty \href
  {http://dx.doi.org/#2} {doi:#2}\else \href {http://dx.doi.org/#2} {#1}\fi
  \endgroup}
\def\mn@eprint#1#2{\mn@eprint@#1:#2::\@nil}
\def\mn@eprint@arXiv#1{\href {http://arxiv.org/abs/#1} {{\tt arXiv:#1}}}
\def\mn@eprint@dblp#1{\href {http://dblp.uni-trier.de/rec/bibtex/#1.xml}
  {dblp:#1}}
\def\mn@eprint@#1:#2:#3:#4\@nil{\def\@tempa {#1}\def\@tempb {#2}\def\@tempc
  {#3}\ifx \@tempc \@empty \let \@tempc \@tempb \let \@tempb \@tempa \fi \ifx
  \@tempb \@empty \def\@tempb {arXiv}\fi \@ifundefined
  {mn@eprint@\@tempb}{\@tempb:\@tempc}{\expandafter \expandafter \csname
  mn@eprint@\@tempb\endcsname \expandafter{\@tempc}}}

\bibitem[\protect\citeauthoryear{Abramowski et~al.}{Abramowski
  et~al.}{2014}]{Abramowski:2014uta}
Abramowski A.,  et~al., 2014, \mn@doi [Astron. Astrophys.]
  {10.1051/0004-6361/201322510}, 562, A145

\bibitem[\protect\citeauthoryear{Ackermann et~al.}{Ackermann
  et~al.}{2018}]{Biteau:2018tmv}
Ackermann M.,  et~al., 2018, \mn@doi [Astrophys. J. Suppl.]
  {10.3847/1538-4365/aacdf7}, 237, 32

\bibitem[\protect\citeauthoryear{Aharonian, Coppi  \& V\"olk}{Aharonian
  et~al.}{1994}]{1994ApJ...423L...5A}
Aharonian F.,  Coppi P.~S.,   V\"olk H.~J.,  1994, \mn@doi [Astrophys. J.
  Lett.] {10.1086/187222}, 423, L5

\bibitem[\protect\citeauthoryear{{Aharonian} et~al.,}{{Aharonian}
  et~al.}{2006a}]{hess2006a}
{Aharonian} F.,  et~al., 2006a, \mn@doi [\nat] {10.1038/nature04680}, \href
  {http://adsabs.harvard.edu/abs/2006Natur.440.1018A} {440, 1018}

\bibitem[\protect\citeauthoryear{Aharonian et~al.}{Aharonian
  et~al.}{2006b}]{Aharonian:2005kn}
Aharonian F.,  et~al., 2006b, \mn@doi [Astrophys. J.] {10.1086/498013}, 636,
  777

\bibitem[\protect\citeauthoryear{{Aliu} et~al.,}{{Aliu}
  et~al.}{2012}]{2012ApJ...755..118A}
{Aliu} E.,  et~al., 2012, \mn@doi [\apj] {10.1088/0004-637X/755/2/118}, \href
  {https://ui.adsabs.harvard.edu/\#abs/2012ApJ...755..118A} {755, 118}

\bibitem[\protect\citeauthoryear{Alves~Batista, Saveliev, Sigl  \&
  Vachaspati}{Alves~Batista et~al.}{2016a}]{AlvesBatista:2016urk}
Alves~Batista R.,  Saveliev A.,  Sigl G.,   Vachaspati T.,  2016a, \mn@doi
  [Phys. Rev. D] {10.1103/PhysRevD.94.083005}, 94, 083005

\bibitem[\protect\citeauthoryear{Alves~Batista et~al.,}{Alves~Batista
  et~al.}{2016b}]{Batista:2016yrx}
Alves~Batista R.,  et~al., 2016b, \mn@doi [J. Cosmol. Astropart. Phys.]
  {10.1088/1475-7516/2016/05/038}, 1605, 038

\bibitem[\protect\citeauthoryear{Arlen, Vassiliev, Weisgarber, Wakely  \&
  Shafi}{Arlen et~al.}{2014}]{Arlen:2012iy}
Arlen T.~C.,  Vassiliev V.~V.,  Weisgarber T.,  Wakely S.~P.,   Shafi S.~Y.,
  2014, \mn@doi [Astrophys. J.] {10.1088/0004-637X/796/1/18}, 796, 18

\bibitem[\protect\citeauthoryear{Atwood et~al.}{Atwood
  et~al.}{2009}]{Atwood:2009ez}
Atwood W.~B.,  et~al., 2009, \mn@doi [Astrophys. J.]
  {10.1088/0004-637X/697/2/1071}, 697, 1071

\bibitem[\protect\citeauthoryear{Bohm \& Gross}{Bohm \&
  Gross}{1949}]{Bohm:1949zz}
Bohm D.,  Gross E.~P.,  1949, \mn@doi [Phys. Rev.] {10.1103/PhysRev.75.1851},
  75, 1851

\bibitem[\protect\citeauthoryear{{Bonnoli}, {Tavecchio}, {Ghisellini}  \&
  {Sbarrato}}{{Bonnoli} et~al.}{2015}]{2015MNRAS.451..611B}
{Bonnoli} G.,  {Tavecchio} F.,  {Ghisellini} G.,   {Sbarrato} T.,  2015,
  \mn@doi [\mnras] {10.1093/mnras/stv953}, \href
  {https://ui.adsabs.harvard.edu/\#abs/2015MNRAS.451..611B} {451, 611}

\bibitem[\protect\citeauthoryear{Brejzman \& Ryutov}{Brejzman \&
  Ryutov}{1974}]{0029-5515-14-6-012}
Brejzman B.~N.,  Ryutov D.~D.,  1974, \mn@doi [Nucl. Fusion]
  {10.1088/0029-5515/14/6/012}, 14, 873

\bibitem[\protect\citeauthoryear{Broderick, Chang  \& Pfrommer}{Broderick
  et~al.}{2012}]{0004-637X-752-1-22}
Broderick A.~E.,  Chang P.,   Pfrommer C.,  2012, \mn@doi [Astrophys. J.]
  {10.1088/0004-637X/752/1/22}, 752, 22

\bibitem[\protect\citeauthoryear{Broderick, Pfrommer, Puchwein  \&
  Chang}{Broderick et~al.}{2014}]{Chang:2013yia}
Broderick A.~E.,  Pfrommer C.,  Puchwein E.,   Chang P.,  2014, \mn@doi
  [Astrophys. J.] {10.1088/0004-637X/790/2/137}, 790, 137

\bibitem[\protect\citeauthoryear{{Burns}, {Skillman}  \& {O'Shea}}{{Burns}
  et~al.}{2010}]{burns2010a}
{Burns} J.~O.,  {Skillman} S.~W.,   {O'Shea} B.~W.,  2010, \mn@doi [\apj]
  {10.1088/0004-637X/721/2/1105}, \href
  {http://adsabs.harvard.edu/abs/2010ApJ...721.1105B} {721, 1105}

\bibitem[\protect\citeauthoryear{{CTA Consortium}}{{CTA
  Consortium}}{2019}]{cta}
{CTA Consortium} 2019, Science with the Cherenkov Telescope Array.
World Scientific, \mn@doi{10.1142/10986}

\bibitem[\protect\citeauthoryear{{Cerruti}, {Zech}, {Boisson}  \&
  {Inoue}}{{Cerruti} et~al.}{2015}]{cerruti2015a}
{Cerruti} M.,  {Zech} A.,  {Boisson} C.,   {Inoue} S.,  2015, \mn@doi [\mnras]
  {10.1093/mnras/stu2691}, \href
  {https://ui.adsabs.harvard.edu/abs/2015MNRAS.448..910C} {448, 910}

\bibitem[\protect\citeauthoryear{Chang, Broderick  \& Pfrommer}{Chang
  et~al.}{2012}]{Chang:2011bf}
Chang P.,  Broderick A.~E.,   Pfrommer C.,  2012, \mn@doi [Astrophys. J.]
  {10.1088/0004-637X/752/1/23}, 752, 23

\bibitem[\protect\citeauthoryear{Chang, Broderick, Pfrommer, Puchwein, Lamberts
   \& Shalaby}{Chang et~al.}{2014}]{Chang:2014cta}
Chang P.,  Broderick A.~E.,  Pfrommer C.,  Puchwein E.,  Lamberts A.,   Shalaby
  M.,  2014, \mn@doi [Astrophys. J.] {10.1088/0004-637X/797/2/110}, 797, 110

\bibitem[\protect\citeauthoryear{Chang, Broderick, Pfrommer, Puchwein,
  Lamberts, Shalaby  \& Vasil}{Chang et~al.}{2016}]{Chang:2016gji}
Chang P.,  Broderick A.~E.,  Pfrommer C.,  Puchwein E.,  Lamberts A.,  Shalaby
  M.,   Vasil G.,  2016, \mn@doi [Astrophys. J.] {10.3847/1538-4357/833/1/118},
  833, 118

\bibitem[\protect\citeauthoryear{Chen, Errando, Buckley  \& Ferrer}{Chen
  et~al.}{2018}]{Chen:2018mjd}
Chen W.,  Errando M.,  Buckley J.~H.,   Ferrer F.,  2018, arXiv

\bibitem[\protect\citeauthoryear{{Costamante}, {Bonnoli}, {Tavecchio},
  {Ghisellini}, {Tagliaferri}  \& {Khangulyan}}{{Costamante}
  et~al.}{2018}]{2018MNRAS.477.4257C}
{Costamante} L.,  {Bonnoli} G.,  {Tavecchio} F.,  {Ghisellini} G.,
  {Tagliaferri} G.,   {Khangulyan} D.,  2018, \mn@doi [\mnras]
  {10.1093/mnras/sty857}, \href
  {https://ui.adsabs.harvard.edu/\#abs/2018MNRAS.477.4257C} {477, 4257}

\bibitem[\protect\citeauthoryear{{Dai}, {Zhang}, {Gou}, {M{\'e}sz{\'a}ros}  \&
  {Waxman}}{{Dai} et~al.}{2002}]{2002ApJ...580L...7D}
{Dai} Z.~G.,  {Zhang} B.,  {Gou} L.~J.,  {M{\'e}sz{\'a}ros} P.,   {Waxman} E.,
  2002, \mn@doi [\apjl] {10.1086/345494}, \href
  {https://ui.adsabs.harvard.edu/abs/2002ApJ...580L...7D} {580, L7}

\bibitem[\protect\citeauthoryear{d'Avezac, Dubus  \& Giebels}{d'Avezac
  et~al.}{2007}]{d'Avezac:2007sg}
d'Avezac P.,  Dubus G.,   Giebels B.,  2007, \mn@doi [Astron. Astrophys.]
  {10.1051/0004-6361:20066712}, 469, 857

\bibitem[\protect\citeauthoryear{{Davidson}, {Hammer}, {Haber}  \&
  {Wagner}}{{Davidson} et~al.}{1972}]{davidson1972a}
{Davidson} R.~C.,  {Hammer} D.~A.,  {Haber} I.,   {Wagner} C.~E.,  1972,
  \mn@doi [Physics of Fluids] {10.1063/1.1693910}, \href
  {http://adsabs.harvard.edu/abs/1972PhFl...15..317D} {15, 317}

\bibitem[\protect\citeauthoryear{Dolag, Kachelrie{\ss}, Ostapchenko  \&
  Tom{\`a}s}{Dolag et~al.}{2009}]{0004-637X-703-1-1078}
Dolag K.,  Kachelrie{\ss} M.,  Ostapchenko S.,   Tom{\`a}s R.,  2009, \mn@doi
  [Astrophys. J.] {10.1088/0004-637X/703/1/1078}, 703, 1078

\bibitem[\protect\citeauthoryear{{Dom{\'\i}nguez} et~al.,}{{Dom{\'\i}nguez}
  et~al.}{2011}]{dominguez2011a}
{Dom{\'\i}nguez} A.,  et~al., 2011, \mn@doi [Mon. Not. R. Astron. Soc.]
  {10.1111/j.1365-2966.2010.17631.x}, 410, 2556

\bibitem[\protect\citeauthoryear{Duplessis \& Vachaspati}{Duplessis \&
  Vachaspati}{2017}]{Duplessis:2017rde}
Duplessis F.,  Vachaspati T.,  2017, \mn@doi [J. Cosmol. Astropart. Phys.]
  {10.1088/1475-7516/2017/05/005}, 1705, 005

\bibitem[\protect\citeauthoryear{Durrer \& Neronov}{Durrer \&
  Neronov}{2013}]{DuNe}
Durrer R.,  Neronov A.,  2013, \mn@doi [Astron. Astrophys. Rev.]
  {10.1007/s00159-013-0062-7}, 21, 62

\bibitem[\protect\citeauthoryear{Essey, Kalashev, Kusenko  \& Beacom}{Essey
  et~al.}{2010}]{Essey:2009ju}
Essey W.,  Kalashev O.~E.,  Kusenko A.,   Beacom J.~F.,  2010, \mn@doi [Phys.
  Rev. Lett.] {10.1103/PhysRevLett.104.141102}, 104, 141102

\bibitem[\protect\citeauthoryear{Essey, Ando  \& Kusenko}{Essey
  et~al.}{2011a}]{Essey:2010nd}
Essey W.,  Ando S.,   Kusenko A.,  2011a, \mn@doi [Astropart. Phys.]
  {10.1016/j.astropartphys.2011.06.010}, 35, 135

\bibitem[\protect\citeauthoryear{Essey, Kalashev, Kusenko  \& Beacom}{Essey
  et~al.}{2011b}]{Essey:2010er}
Essey W.,  Kalashev O.,  Kusenko A.,   Beacom J.~F.,  2011b, \mn@doi
  [Astrophys. J.] {10.1088/0004-637X/731/1/51}, 731, 51

\bibitem[\protect\citeauthoryear{Fan, Dai  \& Wei}{Fan
  et~al.}{2004}]{Fan:2003qr}
Fan Y.-Z.,  Dai Z.~G.,   Wei D.~M.,  2004, \mn@doi [Astron. Astrophys.]
  {10.1051/0004-6361:20034472}, 415, 483

\bibitem[\protect\citeauthoryear{Finke, Razzaque  \& Dermer}{Finke
  et~al.}{2010}]{0004-637X-712-1-238}
Finke J.~D.,  Razzaque S.,   Dermer C.~D.,  2010, \mn@doi [Astrophys. J.]
  {10.1088/0004-637X/712/1/238}, 712, 238

\bibitem[\protect\citeauthoryear{Fitoussi, Belmont, Malzac, Marcowith,
  Cohen-Tanugi  \& Jean}{Fitoussi et~al.}{2017}]{Fitoussi:2017ble}
Fitoussi T.,  Belmont R.,  Malzac J.,  Marcowith A.,  Cohen-Tanugi J.,   Jean
  P.,  2017, \mn@doi [Mon. Not. R. Astron. Soc.] {10.1093/mnras/stw3365}, 466,
  3472

\bibitem[\protect\citeauthoryear{Franceschini, Rodighiero  \&
  Vaccari}{Franceschini et~al.}{2008}]{Franceschini:2008tp}
Franceschini A.,  Rodighiero G.,   Vaccari M.,  2008, \mn@doi [Astron.
  Astrophys.] {10.1051/0004-6361:200809691}, 487, 837

\bibitem[\protect\citeauthoryear{Fried}{Fried}{1959}]{1959PhFl....2..337F}
Fried B.~D.,  1959, \mn@doi [Phys. Fluids] {10.1063/1.1705933}, 2, 337

\bibitem[\protect\citeauthoryear{Galeev, Sagdeev, Sigov, Shapiro  \&
  Shevchenko}{Galeev et~al.}{1975}]{1975FizPl...1...10G}
Galeev A.~A.,  Sagdeev R.~Z.,  Sigov I.~S.,  Shapiro V.~D.,   Shevchenko V.~I.,
   1975, Sov. J. Plasma Phys., 1, 5

\bibitem[\protect\citeauthoryear{Galeev, Sagdeev, Shapiro  \&
  Shevchenko}{Galeev et~al.}{1977}]{GaleevJETP1977}
Galeev A.~A.,  Sagdeev R.~Z.,  Shapiro V.~D.,   Shevchenko V.~I.,  1977, Zh.
  Eksp. Teor. Fiz. [Sov. Phys. JETP], 45, 266

\bibitem[\protect\citeauthoryear{Gehrels \& Meszaros}{Gehrels \&
  Meszaros}{2012}]{Gehrels:2012kp}
Gehrels N.,  Meszaros P.,  2012, \mn@doi [Science] {10.1126/science.1216793},
  337, 932

\bibitem[\protect\citeauthoryear{Gilmore, Somerville, Primack  \&
  Dom{\`i}nguez}{Gilmore et~al.}{2012}]{Gilmore:2011ks}
Gilmore R.~C.,  Somerville R.~S.,  Primack J.~R.,   Dom{\`i}nguez A.,  2012,
  \mn@doi [Mon. Not. R. Astron. Soc.] {10.1111/j.1365-2966.2012.20841.x}, 422,
  3189

\bibitem[\protect\citeauthoryear{Grognard}{Grognard}{1975}]{Grognard1975}
Grognard R. J.-M.,  1975, \mn@doi [Aust. J. Phys.] {10.1071/PH750731}, 28, 731

\bibitem[\protect\citeauthoryear{{HESS Collaboration} et~al.,}{{HESS
  Collaboration} et~al.}{2013}]{2013MNRAS.434.1889H}
{HESS Collaboration} et~al., 2013, \mn@doi [\mnras] {10.1093/mnras/stt1081},
  \href {https://ui.adsabs.harvard.edu/\#abs/2013MNRAS.434.1889H} {434, 1889}

\bibitem[\protect\citeauthoryear{Hauser \& Dwek}{Hauser \&
  Dwek}{2001}]{Ann.Rev.Astr.Astroph.39.1.249}
Hauser M.~G.,  Dwek E.,  2001, \mn@doi [Ann. Rev. Astron. Astrophys.]
  {10.1146/annurev.astro.39.1.249}, 39, 249

\bibitem[\protect\citeauthoryear{Heiter, Kuempel, Walz  \& Erdmann}{Heiter
  et~al.}{2018}]{Heiter:2017cev}
Heiter C.,  Kuempel D.,  Walz D.,   Erdmann M.,  2018, \mn@doi [Astropart.
  Phys.] {10.1016/j.astropartphys.2018.05.003}, 102, 39

\bibitem[\protect\citeauthoryear{Hofmann}{Hofmann}{2000}]{Hofmann:1999ew}
Hofmann W.,  2000, \mn@doi [AIP Conf. Proc.] {10.1063/1.1291416}, 515, 500

\bibitem[\protect\citeauthoryear{Hui \& Gnedin}{Hui \&
  Gnedin}{1997}]{Hui:1997dp}
Hui L.,  Gnedin N.~Y.,  1997, \mn@doi [Mon. Not. R. Astron. Soc.]
  {10.1093/mnras/292.1.27}, 292, 27

\bibitem[\protect\citeauthoryear{{Huntington} et~al.,}{{Huntington}
  et~al.}{2015}]{2015NatPh..11..173H}
{Huntington} C.~M.,  et~al., 2015, \mn@doi [Nature Physics]
  {10.1038/nphys3178}, \href
  {https://ui.adsabs.harvard.edu/\#abs/2015NatPh..11..173H} {11, 173}

\bibitem[\protect\citeauthoryear{Kaplan \& Tsytovich}{Kaplan \&
  Tsytovich}{1968}]{1968SvA....11..956K}
Kaplan S.~A.,  Tsytovich V.~N.,  1968, Sov. Astron., 11, 956

\bibitem[\protect\citeauthoryear{Kneiske \& Dole}{Kneiske \&
  Dole}{2010}]{2010A&A...515A..19K}
Kneiske T.~M.,  Dole H.,  2010, \mn@doi [Astron. Astrophys.]
  {10.1051/0004-6361/200912000}, 515, A19

\bibitem[\protect\citeauthoryear{Kneiske, Bretz, Mannheim  \& Hartmann}{Kneiske
  et~al.}{2004}]{Kneiske2004}
Kneiske T.~M.,  Bretz T.,  Mannheim K.,   Hartmann D.~H.,  2004, \mn@doi
  [Astron. Astrophys.] {10.1051/0004-6361:20031542}, 413, 807

\bibitem[\protect\citeauthoryear{{McDonald}, {Miralda-Escud{\'e}}, {Rauch},
  {Sargent}, {Barlow}  \& {Cen}}{{McDonald} et~al.}{2001}]{2001ApJ...562...52M}
{McDonald} P.,  {Miralda-Escud{\'e}} J.,  {Rauch} M.,  {Sargent} W. L.~W.,
  {Barlow} T.~A.,   {Cen} R.,  2001, \mn@doi [\apj] {10.1086/323426}, \href
  {https://ui.adsabs.harvard.edu/\#abs/2001ApJ...562...52M} {562, 52}

\bibitem[\protect\citeauthoryear{Medvedev \& Loeb}{Medvedev \&
  Loeb}{1999}]{Medvedev:1999tu}
Medvedev M.~V.,  Loeb A.,  1999, \mn@doi [Astrophys. J.] {10.1086/308038}, 526,
  697

\bibitem[\protect\citeauthoryear{Miniati \& Elyiv}{Miniati \&
  Elyiv}{2013}]{Miniati:2012ge}
Miniati F.,  Elyiv A.,  2013, \mn@doi [Astrophys. J.]
  {10.1088/0004-637X/770/1/54}, 770, 54

\bibitem[\protect\citeauthoryear{Murase, Takahashi, Inoue, Ichiki  \&
  Nagataki}{Murase et~al.}{2008}]{1538-4357-686-2-L67}
Murase K.,  Takahashi K.,  Inoue S.,  Ichiki K.,   Nagataki S.,  2008, \mn@doi
  [Astrophys. J. Lett.] {10.1086/592997}, 686, L67

\bibitem[\protect\citeauthoryear{Nakar, Bret  \& Milosavjevic}{Nakar
  et~al.}{2011}]{Nakar:2011mt}
Nakar E.,  Bret A.,   Milosavjevic M.,  2011, \mn@doi [Astrophys. J.]
  {10.1088/0004-637X/738/1/93}, 738, 93

\bibitem[\protect\citeauthoryear{Neronov \& Semikoz}{Neronov \&
  Semikoz}{2007}]{JETPLett.85.10.473}
Neronov A.,  Semikoz D.~V.,  2007, \mn@doi [JETP Lett.]
  {10.1134/S0021364007100013}, 85, 473

\bibitem[\protect\citeauthoryear{Neronov \& Semikoz}{Neronov \&
  Semikoz}{2009}]{PhysRevD.80.123012}
Neronov A.,  Semikoz D.~V.,  2009, \mn@doi [Phys. Rev. D]
  {10.1103/PhysRevD.80.123012}, 80, 123012

\bibitem[\protect\citeauthoryear{Neronov \& Vovk}{Neronov \&
  Vovk}{2010}]{Neronov02042010}
Neronov A.,  Vovk I.,  2010, \mn@doi [Science] {10.1126/science.1184192}, 328,
  73

\bibitem[\protect\citeauthoryear{Papadopoulos}{Papadopoulos}{1975}]{1975PhFl...18.1769P}
Papadopoulos K.,  1975, \mn@doi [Phys. Fluids] {10.1063/1.861096}, 18, 1769

\bibitem[\protect\citeauthoryear{Pavan, Yoon  \& Umeda}{Pavan
  et~al.}{2011}]{2011PhPl...18d2307P}
Pavan J.,  Yoon P.~H.,   Umeda T.,  2011, \mn@doi [Phys. Plasmas]
  {10.1063/1.3574359}, 18, 042307

\bibitem[\protect\citeauthoryear{{Perucho}, {Quilis}  \&
  {Mart{\'\i}}}{{Perucho} et~al.}{2011}]{perucho2011a}
{Perucho} M.,  {Quilis} V.,   {Mart{\'\i}} J.-M.,  2011, \mn@doi [\apj]
  {10.1088/0004-637X/743/1/42}, \href
  {https://ui.adsabs.harvard.edu/abs/2011ApJ...743...42P} {743, 42}

\bibitem[\protect\citeauthoryear{{Pfrommer}, {Chang}  \&
  {Broderick}}{{Pfrommer} et~al.}{2012}]{2012ApJ...752...24P}
{Pfrommer} C.,  {Chang} P.,   {Broderick} A.~E.,  2012, \mn@doi [\apj]
  {10.1088/0004-637X/752/1/24}, \href
  {https://ui.adsabs.harvard.edu/\#abs/2012ApJ...752...24P} {752, 24}

\bibitem[\protect\citeauthoryear{Plaga}{Plaga}{1995}]{Plaga1994}
Plaga R.,  1995, \mn@doi [Nature] {10.1038/374430a0}, 374, 430

\bibitem[\protect\citeauthoryear{Protheroe \& Biermann}{Protheroe \&
  Biermann}{1996}]{Protheroe:1996si}
Protheroe R.~J.,  Biermann P.~L.,  1996, \mn@doi [Astropart. Phys.]
  {10.1016/S0927-6505(96)00041-2}, 6, 45

\bibitem[\protect\citeauthoryear{Quinn et~al.,}{Quinn
  et~al.}{1996}]{1996ApJ...456L..83Q}
Quinn J.,  et~al., 1996, \mn@doi [Astrophys. J. Lett.] {10.1086/309878}, 456,
  L83

\bibitem[\protect\citeauthoryear{Reimer \& B\"ottcher}{Reimer \&
  B\"ottcher}{2013}]{Reimer:2012vw}
Reimer A.,  B\"ottcher M.,  2013, \mn@doi [Astrophys. J.]
  {10.1016/j.astropartphys.2012.05.011}, 43, 103

\bibitem[\protect\citeauthoryear{Rico}{Rico}{2017}]{Rico:2017euq}
Rico J.,  2017, \mn@doi [AIP Conf. Proc.] {10.1063/1.4968984}, 1792, 060001

\bibitem[\protect\citeauthoryear{Schamel, Lee  \& Morales}{Schamel
  et~al.}{1976}]{1976PhFl...19..849S}
Schamel H.,  Lee Y.~C.,   Morales G.~J.,  1976, \mn@doi [Phys. Fluids]
  {10.1063/1.861550}, 19, 849

\bibitem[\protect\citeauthoryear{Schlickeiser, Vainio, Boettcher, Lerche, Pohl
  \& Schuster}{Schlickeiser et~al.}{2002}]{Schlickeiser:2002dt}
Schlickeiser R.,  Vainio R.,  Boettcher M.,  Lerche I.,  Pohl M.,   Schuster
  C.,  2002, \mn@doi [Astron. Astrophys.] {10.1051/0004-6361:20020975}, 393, 69

\bibitem[\protect\citeauthoryear{Schlickeiser, Ibscher  \& Supsar}{Schlickeiser
  et~al.}{2012}]{0004-637X-758-2-102}
Schlickeiser R.,  Ibscher D.,   Supsar M.,  2012, \mn@doi [Astrophys. J.]
  {10.1088/0004-637X/758/2/102}, 758, 102

\bibitem[\protect\citeauthoryear{Schlickeiser, Krakau  \& Supsar}{Schlickeiser
  et~al.}{2013}]{Schlickeiser:2013eca}
Schlickeiser R.,  Krakau S.,   Supsar M.,  2013, \mn@doi [Astrophys. J.]
  {10.1088/0004-637X/777/1/49}, 777, 49

\bibitem[\protect\citeauthoryear{Shalaby, Broderick, Chang, Pfrommer, Lamberts
  \& Puchwein}{Shalaby et~al.}{2017}]{Shalaby:2017kpr}
Shalaby M.,  Broderick A.~E.,  Chang P.,  Pfrommer C.,  Lamberts A.,   Puchwein
  E.,  2017, \mn@doi [Astrophys. J.] {10.3847/1538-4357/aa8b17}, 848, 81

\bibitem[\protect\citeauthoryear{Shalaby, Broderick, Chang, Pfrommer, Lamberts
  \& Puchwein}{Shalaby et~al.}{2018}]{Shalaby:2018jja}
Shalaby M.,  Broderick A.~E.,  Chang P.,  Pfrommer C.,  Lamberts A.,   Puchwein
  E.,  2018, \mn@doi [Astrophys. J.] {10.3847/1538-4357/aabe92}, 859, 45

\bibitem[\protect\citeauthoryear{Sironi \& Giannios}{Sironi \&
  Giannios}{2014}]{Sironi:2013qfa}
Sironi L.,  Giannios D.,  2014, \mn@doi [Astrophys. J.]
  {10.1088/0004-637X/787/1/49}, 787, 49

\bibitem[\protect\citeauthoryear{Stecker, Malkan  \& Scully}{Stecker
  et~al.}{2006}]{Stecker:2005qs}
Stecker F.~W.,  Malkan M.~A.,   Scully S.~T.,  2006, \mn@doi [Astrophys. J.]
  {10.1086/506188}, 648, 774

\bibitem[\protect\citeauthoryear{{Stecker}, {Scully}  \& {Malkan}}{{Stecker}
  et~al.}{2016}]{stecker2016a}
{Stecker} F.~W.,  {Scully} S.~T.,   {Malkan} M.~A.,  2016, \mn@doi [Astrophys.
  J.] {10.3847/0004-637X/827/1/6}, 827, 6

\bibitem[\protect\citeauthoryear{Strong, Moskalenko  \& Reimer}{Strong
  et~al.}{2004}]{Strong:2004de}
Strong A.~W.,  Moskalenko I.~V.,   Reimer O.,  2004, \mn@doi [Astrophys. J.]
  {10.1086/423193}, 613, 962

\bibitem[\protect\citeauthoryear{Tashiro, Chen, Ferrer  \& Vachaspati}{Tashiro
  et~al.}{2014}]{Tashiro:2013ita}
Tashiro H.,  Chen W.,  Ferrer F.,   Vachaspati T.,  2014, \mn@doi [Mon. Not. R.
  Astron. Soc.] {10.1093/mnrasl/slu134}, 445, L41

\bibitem[\protect\citeauthoryear{Taylor, Vovk  \& Neronov}{Taylor
  et~al.}{2011}]{Taylor:2011bn}
Taylor A.~M.,  Vovk I.,   Neronov A.,  2011, \mn@doi [Astron. Astrophys.]
  {10.1051/0004-6361/201116441}, 529, A144

\bibitem[\protect\citeauthoryear{Vafin, Rafighi, Pohl  \& Niemiec}{Vafin
  et~al.}{2018}]{Vafin:2018kox}
Vafin S.,  Rafighi I.,  Pohl M.,   Niemiec J.,  2018, \mn@doi [Astrophys. J.]
  {10.3847/1538-4357/aab552}, 857, 43

\bibitem[\protect\citeauthoryear{Vovk, Taylor, Semikoz  \& Neronov}{Vovk
  et~al.}{2012}]{Vovk:2011aa}
Vovk I.,  Taylor A.~M.,  Semikoz D.,   Neronov A.,  2012, \mn@doi [Astrophys.
  J. Lett.] {10.1088/2041-8205/747/1/L14}, 747, L14

\bibitem[\protect\citeauthoryear{Weekes et~al.,}{Weekes
  et~al.}{1989}]{Weekes:1989tc}
Weekes T.~C.,  et~al., 1989, \mn@doi [Astrophys. J.] {10.1086/167599}, 342, 379

\bibitem[\protect\citeauthoryear{Weekes et~al.}{Weekes
  et~al.}{2002}]{Weekes:2001pd}
Weekes T.~C.,  et~al., 2002, \mn@doi [Astrophys. J.]
  {10.1016/S0927-6505(01)00152-9}, 17, 221

\bibitem[\protect\citeauthoryear{Weibel}{Weibel}{1959}]{Weibel:1959zz}
Weibel E.~S.,  1959, \mn@doi [Phys. Rev. Lett.] {10.1103/PhysRevLett.2.83}, 2,
  83

\bibitem[\protect\citeauthoryear{Yan, Zhou, Zhang, Zhu  \& Wang}{Yan
  et~al.}{2019}]{Yan:2018pca}
Yan D.,  Zhou J.,  Zhang P.,  Zhu Q.,   Wang J.,  2019, \mn@doi [Astrophys. J.]
  {10.3847/1538-4357/aaef7d}, 870, 17

\bibitem[\protect\citeauthoryear{Yang \& Dai}{Yang \& Dai}{2015}]{Yang:2015lqy}
Yang Y.-P.,  Dai Z.-G.,  2015, \mn@doi [Res. Astron. Astrophys.]
  {10.1088/1674-4527/15/12/005}, 15, 2173

\makeatother
\end{thebibliography}





\bsp	
\label{lastpage}
\end{document}